\documentclass[]{iucr}              

\usepackage[]{graphicx}
\usepackage{amsmath}
\usepackage{amsfonts}
\usepackage{textcomp}
\usepackage[abs]{overpic}

\newcommand{\specialcell}[2][c]{
  \begin{tabular}[#1]{@{}c@{}}#2\end{tabular}}

\usepackage{color}
\usepackage[usenames,dvipsnames,svgnames,table]{xcolor}
\usepackage{enumerate}

     \journalcode{}              

\begin{document}                  



\title{Simulation of X-ray diffraction profiles for bent anisotropic crystals}


\cauthor[a]{Manuel}{Sanchez del Rio}{srio@esrf.eu}{address if different from \aff}
\author[a]{Nicolas}{Perez-Bocanegra}
\author[a]{Xianbo}{Shi}
\author[a]{Veijo}{Honkim{\"{a}}ki}
\author[a]{Lin}{Zhang}

\aff[a]{ ESRF - The European Synchrotron, 71 Avenue des Martyrs, 38000 Grenoble \country{France}}









\maketitle                        

\begin{synopsis}
The equations for simulating diffraction profiles and their implementation in computer code are presented for 
the cases of meridional and sagittal bent crystals using the multilamellar and Penning-Polder theories and
including crystal anisotropy. 

\end{synopsis}

\begin{abstract}
The equations for calculating diffraction profiles for bent crystals are revisited for both meridional and sagittal
bending. Two approximated methods for computing diffraction profiles 
are treated: multilamellar and Penning-Polder. A common treatment of crystal anisotropy is included in these models.
The formulation presented is implemented into the XOP package, completing and
updating the crystal module {\color{black} that simulates diffraction profiles for perfect, mosaic and now distorted 
crystals by elastic bending}.  
\end{abstract}



\section{Introduction}

The motivation of this work is the availability of easy-to-use computer codes for simulating diffraction or 
reflection profiles of bent crystals. The target audience is the users and researchers of 
synchrotron radiation facilities, but also other fields of X-ray research, like plasma physics. 
There is a vast literature on the theory and applicability
of the Dynamical Theory of Diffraction to the calculation of diffraction profiles (see e.g, \cite{Authier} for an
updated review). However, many scientists and engineers need to know the performances of bent crystals under X-rays 
without becoming specialist in the field of Dynamical Diffraction. 
Several available codes may be used, many of them available via collaborations with their authors: 
REFLECT \cite{REFLECT}, REFLEX \cite{Caciuffo}, PEPO \cite{Schulze}, DIXI \cite{DIXI}, and others publically 
available \cite{Stepanov}.
One popular code for the synchrotron community is XOP \cite{Xop2011}, a graphical environment for computer codes 
for i) modeling of X-ray sources (e.g., synchrotron radiation sources, such as undulators and wigglers), 
ii) calculating characteristics of optical elements (mirrors, filters, crystals, multilayers, etc.), and 
iii) multipurpose data visualizations and analyses. XOP is used extensively to simulate crystal diffraction 
profiles for perfect, bent, and mosaic crystals. 

The calculation of diffraction profiles for flat (undistorted) crystals is usually performed using the basic 
equations of the Dynamical Theory, in different formulations. Although a full quantum treatment of X-ray 
diffraction by a crystal exists \cite{ashkin,kuriyama}, a {\it classical} approach, as discussed in 
\cite{hartwig_hierarchy} is adopted for most practical cases. In the classical approach, the propagation of 
the X-ray field inside and outside the crystal is described by the Maxwell equations assuming that all quantities
(electric susceptibility, electric field, etc.) are defined in a continuous way for any point in the space and time,
and the interaction of the X-ray field with the crystal electrons is described by quantum mechanics. 
The dynamical theories for X-ray diffraction in perfect crystals have been extended to include lattice distortions 
related to crystalline defects and macroscopic bending. For practical purposes, several {\it levels of 
approximations} may be defined \cite{hartwig_hierarchy}: 

\begin{enumerate}[(a)]
 \item dynamical diffraction theory for the perfect crystal,
 \item local applications of the dynamical theory for perfect crystals to distorted ones,
 \item geometrical optics or eikonal theory, and
 \item wave optics or Takagi theory.
\end{enumerate}

{\color{black} 
Item (a) is related to the classical dynamical diffraction of X-rays in perfect crystals. 
The term dynamical is used when the rescattering and absorption of the X-rays in the crystal volume 
is considered. Perfect crystals are ideal 
undistorted monocrystals over a large distances (as compared with the unit cells) thus assuming perfect 
alignment of the atoms in the crystalline structure. There is no curvature of the atomic planes 
(like the originated by elastic bending or thermal distortion) and no alteration of the atomic order (no inclusions,
dislocations, defects, stress, cracks, etc). Since the pioneering 
work of \cite{Darwin14}, several several formulations are available \cite{Ewald17}, \cite{Laue31}, \cite{Zachariasen}, etc. 
Some comprehensive books describe them \cite{James}, \cite{Pinsker} and \cite{Authier}, which includes a complete historical review.

The (b) approximation works for crystals with small distortions with respect to the perfect crystals. This means that the 
displacement vector $\vec{u}(\vec{r})$ varies {\it very slowly}. The crystal deformation transforms a point with 
position $\vec{r}$ into another at $\vec{r'}$, with $\vec{u}(\vec{r})=\vec{r'}-\vec{r}\approx\vec{u}(\vec{r'} )$). 
The crystal reflectivity for diffracted beam can therefore be computed quantitatively 
by shifting the angular position of the incident direction by a value equal to the effective misorientation,
and applying the diffracted intensity of the perfect crystal. 
This method was introduced by \cite{Bonse} and \cite{Authier1966}. Also, the multilamellar method
\cite{Caciuffo,Erola} discussed later belongs to this category. It can be used for estimating the diffraction 
profile in both Bragg and Laue geometries in many practical cases of crystal curvature.

The diffraction theory of the (c) approximation mimics the geometrical optics for visible light. An inductive 
derivation was made by \cite{Penning}, a deductive theory was presented in \cite{Kato1,Kato2,Kato3}, and a derivation 
from the Takagi-Taupin Equations is in \cite{IndenbomLevel3}. Level (d) rely on the Takagi-Taupin equations 
\cite{Takagi,Taupin}.
}

For calculations of perfect undistorted crystals XOP implements the equations of the dynamical theory of diffraction from
\cite{Zachariasen}, summarized in Section \ref{undistorted}. On the other hand, imperfect mosaic crystals can be simulated in XOP 
also using Zachariasen theory which is valid for mosaicity values much larger than the Darwin width. 

In this work we extended the XCRYSTAL application in XOP to cover perfect crystals with small deformations originated
by elastic bending. {\color{black} Elastic bending produces a crystal distortion that depends on its anisotropy. 
Every crystalline material is anisotropic, meaning the the elastic constants are not scalar, but depend on the direction
(they are deduced from the compliance or stiffness tensors). Anisotropy is important when the crystal is bent, and is not relevant 
for the perfect undistorted crystals. We distinguish curvature in two directions with respect to the direction of the incident beam: 
meridional curvature in a plane that contains the incident direction, and sagittal, a plane perpendicular to it. The meridional plane
coincides with the diffraction plane. A cylindrically bent crystal is curved in one single plane. A single moment is sufficient to bent the 
crystal in one direction. Two moments along perpendicular planes will bend the crystal in two directions. 
Perfect cylinders are difficult to obtain by elastic bending, because when applying one-moment to a crystal block or plate there is 
an spurious curvature in the plane perpendicular to the main bending plane (anticlastic curvature). 
Spherical or toroidal crystals are curved in both planes, usually made by applying two moments. 
The elastic constants and tensors are related to the principal crystallographic directions, which are coincident with the crystal 
block directions only if the crystals is {\it symmetric}. i.e., the crystalline planes are parallel to the crystal faces. In the most general case, the crystal planes are not parallel to the 
crystallographic directions and the crystal is called {\it asymmetric}. For the diffraction effects, a crystal curved with a non constant 
radius of curvature (parabolic, ellipsoidal, conic, etc.) can be approximated as a crystal with two averaged curvatures over the meridional and 
sagittal planes.
}

For simulating the diffraction profiles, two approximated theories are used: the multilamellar method and the Penning-Polder theory. 

The multilamellar method described in Section \ref{multilamellar} can be used to simulate
diffraction profiles of curved crystals in both Bragg and Laue geometries. This models belongs to
level of approximation (b). We develop the formulation of the multilamellar theory working for both sagittal and meridional curvatures in 
both Bragg and Laue geometries. This requires a correct treatment of the crystal anisotropy. The unified
formulation presented here extends the cases treated in literature for isotropic Bragg crystals like 
\cite{Caciuffo,Erola} and for anisotropic Laue crystals \cite{xianbo_spie} to the general case of two-moment
bending Laue or Bragg anisotropic crystals. 

The Penning-Polder model \cite{Penning} summarized in Section \ref{penningpolder} belongs to 
the level of approximations (c), and only applies to Laue geometry. 
This model has been successfully applied for calculating the diffraction profiles of high energy monochromators used in 
beamlines at many synchrotrons (APS, NSLS, ESRF, Spring-8, Petra, Diamond, etc.). 
Again, we present here a formulation that can be applied for crystals bent in two directions (meridional and sagittaly, 
thus unifying previous uses of the Penning-Polder theory for meridionally bent crystals, isotropic 
\cite{Sanchez1997} or anisotropic \cite{Schulze}, or for sagittal bending \cite{xianbo_spie}. 
Approximation level (d), thus solving Takagi-Taupin equations will be addressed in a future work. 

From the computer point of view, a unification of and modernization of the XOP crystal module has been done in 
order to upgrade, clean and improve its structure. A single Fortran 95 module calculates now crystal diffraction 
using the different calculation algorithms described here, using a full 3D vectorial calculus for beam 
direction and crystal orientation. Thus, this paper is a good companion of the software package and will be 
used as reference manual. It is designed for being integrated in other X-ray codes, like for the ray tracing 
package SHADOW \cite{SHADOW}. 

\section{Algorithms for computing diffraction by bent crystals}

Using a crystal reflection defined by the Miller indices $hkl$, the reciprocal vector of the lattice is 
$\vec{H} = (1/d_{hkl}) \vec{n}^H$, with $d_{hkl}$ the interplanar distance, and $\vec{n}^H$ a unitary vector normal to 
the Bragg planes [$hkl$]. 
 
The Laue equation 
\begin{equation}
 \label{eq:laue}
 \vec{k}^H_B = \vec{k}^0_B + \vec{H},
\end{equation}
which is satisfied only for the diffraction condition, gives the wavevector of the diffracted wave $\vec{k}^H_B$ for a particular
position of the incident wavevector $\vec{k}^0_B$, that is, its angle with the reflecting $[hkl]$ planes is 
the Bragg angle $\theta_B$. This gives the Bragg law $\lambda=2 d_{hkl} \sin\theta_B$, with $\lambda$ the photon 
wavelength. In this paper $|\vec{k}|=1/\lambda$ as in the text of \cite{Zachariasen}. 

The change in the direction of any {\it monochromatic beam} (not necessarily satisfying the diffraction condition or Laue equation)
diffracted by a crystal (Laue or Bragg) can be calculated using i) elastic scattering in the diffraction process:
\begin{equation}
   |\vec{k^{0}}| = |\vec{k^{H}}|=\frac{1}{\lambda},
\end{equation}
with $\vec{k}^{0,H}=(1/\lambda)\vec{V}^{0,H}$, and $\vec{V}$ a unitary vector; 
and ii) the boundary conditions at the crystal surface:
\begin{equation}
 \label{eq:parallel_cte}
 \vec{k}^H_{||} = \vec{k}^0_{||} + \vec{H}_{||},
\end{equation}
where $||$ refers to the component parallel to the crystal surface.

A crystal cut is defined by $\vec{n}$, a unity vector normal to the crystal surface pointing outside the crystal bulk. 
Usually it is expressed as a function of $\alpha$ in Bragg geometry, and   $\chi$ in Laue geometry, but if both values are 
well-defined they can be used indistinctly in both geometries (see Appendix~\ref{preliminaries}).
The projections of the beam directions onto this vector are: $\gamma_0=\vec{n}~\cdot~\vec{V^0}$ and $\gamma_H=\vec{n}~\cdot~\vec{V^H}$. 
The asymmetry factor is $b = (\vec{n} \cdot \vec{k^0})/(\vec{n} \cdot (\vec{k^0}+\vec{H}) ) \approx \gamma_0/\gamma_H$.

\subsection{The perfect crystal}
\label{undistorted}

For the perfect (undistorted) crystal we follow the~\cite{Zachariasen} formulation, because of its accuracy, compactness and easy 
numerical implementation. This formulation expresses the crystal reflectivity as a function of angular parameter: 
       \begin{equation}
        \label{alphaZac}
	\alpha_Z = \frac{1}{|\vec{k}^0|^2} \left[ |\vec{H}|^2 + 2 \vec{k}^0~.~\vec{H} \right]
       \end{equation}
which measures the separation of the incident field from the Bragg condition either in angular terms
(``rotating crystal'', for a fixed photon wavelength): 
\begin{equation}
\label{angle_scan}
\alpha_Z \approx 2(\theta_B - \theta) \sin(2 \theta_B) = -2 \Delta \theta \sin(2 \theta_B), 
\end{equation}
or in terms of photon wavelength (or energy): 
(``Laue method'')
\begin{equation}
\label{energy_scan}
\alpha_Z \approx 4 \frac{\lambda-\lambda_B}{\lambda_B} \sin^2 \theta_B = 4 \frac{E_B-E}{E_B} \sin^2 \theta_B, 
\end{equation}
However, it is recommended to compute $\alpha_Z$ using the exact expression in Eq. (\ref{alphaZac}) which is also valid in 
extreme cases, like in normal incidence. 

The X-ray reflectivity of a single perfect parallel-sided crystal in Bragg (or reflection) geometry is:

	\begin{equation}
	\label{ZacBragg}
	r^{bragg}(\alpha_Z) \equiv \frac{1}{|b|}  \frac{I^H}{I^0}  =
	\frac{1}{|b|} 
	\left| \frac{x_1 x_2 (c_1 - c_2)}{c_2 x_2-c_1 x_1} \right|^{2}.
	\end{equation}
For Laue (or transmission) geometry we have:
	\begin{equation}
	\label{ZacLaue}
	r^{laue}(\alpha_Z) \equiv \frac{1}{|b|}  \frac{I^H}{I^0}  =
	\frac{1}{|b|} 
	\left| \frac{x_1 x_2 (c_1 - c_2)}{x_2-x_1} \right|^{2}.
	\end{equation}
The transmitivity (forward diffracted beam) are  
	\begin{equation}
	\label{ZacTransBragg}
	t^{bragg}(\alpha_Z) =
	\left| \frac{c_1 c_2 (x_2 - x_1)}{c_2 x_2-c_1 x_1} \right|^{2},
	\end{equation}

	\begin{equation}
	\label{ZacTransLaue}
	t^{laue}(\alpha_Z) =
	\left| \frac{x_2 c_1 - x_1 c_2}{x_2-x_1} \right|^{2},
	\end{equation}
$I^0$ is the intensity of the incident wave along direction $\vec{V^{0}}$ with wavevector $\vec{k^0}$ ,
$I^H$ is the intensity of the external diffracted wave along the direction $\vec{V^{H}}$ with wavevector $\vec{k^H}$ ,
$c_{1,2}$ are phase terms dependent on the crystal thickness $T$, and both $x_{1,2}$ and $c_{1,2}$ terms depend on 
$\alpha_Z$ and on the crystal electrical susceptibility in a rather non-trivial way:   

\begin{eqnarray}
\label{x_zac}
\left( \begin{array}{ll}
               x_{1} \\
               x_{2}
       \end{array} 
\right)
= \frac{- z\pm \sqrt{qP^2+z^2}}{P\Psi_{\bar{H}}}, \nonumber \\
z = \frac{1-b}{2} \Psi_0 + \frac{b}{2} \alpha_Z, 
\end{eqnarray}
$q=b\Psi_H\Psi_{\bar{H}}$,
$P$  is the polarization factor ($P=1$ for $\sigma$-polarization, $P=|\cos2\theta_B|$ for $\pi$-polarization),
and $\Psi_H$ is the Fourier component of the electrical susceptibility $\Psi_0$ related to the structure factor
$F_H$ as: 
\begin{equation}
 \Psi_H = \frac{-r_0 \lambda^2}{\pi v_c} F_H;~~~ r_0=\frac{e^2}{mc^2},
\end{equation}
where $r_0$ the classical electron radius, $v_c$ the volume of the unit cell, $e$ the charge of the electron and 
$c$ the speed of light. 

The $c_{1,2}$ phases in Eqs.~\ref{ZacBragg}-\ref{ZacTransLaue} are expressed as:

\begin{eqnarray}
\label{cZachariasen}
c_1=e^{-i \phi_1 T} \nonumber \\
c_2 =e^{-i \phi_2 T} \nonumber \\
\phi_1 = -\frac{2\pi k^0 \delta'_0}{\gamma_0} \\
\phi_2 = -\frac{2\pi k^0 \delta''_0}{\gamma_0} \nonumber
\end{eqnarray}

and the other quantities are defined as:
       \begin{eqnarray}
       \left( \begin{array}{ll}
               \delta_0' \\
               \delta_0''
	       \end{array} 
	\right)
	= \frac{1}{2} \left( \Psi_0 - z\pm \sqrt{qP^2+z^2} \right) , \nonumber \\
	\end{eqnarray}

Note that a factor $|b|^{-1}$ appears for calculating the diffracted beams 
(Eqs. (\ref{ZacBragg}) and (\ref{ZacLaue})) to guarantee the conservation of the total power when the linear width
of the incident beam is large compared with the depth of penetration in the crystal, as discussed in 
\cite{Zachariasen} (pag. 122). This factor is not present for the transmitted beams
(Eqs. (\ref{ZacTransBragg}) and (\ref{ZacTransLaue})). 
Also note that the signs of $\phi_{1,2}$ in Eq. \ref{cZachariasen} are changed with respect to \cite{Zachariasen} because in our 
definitions the surface normal points outside the crystal.

$\alpha_Z$ in Eq. (\ref{alphaZac}) is the magnitude that measures the separation of the incident beam $\vec{k}^0$ from the 
Bragg position. One can define a dimensionless parameter $\eta$ that measures the ``normalized'' angular 
separation or ``deviation parameter''
\cite{Zachariasen},\cite{Authier}.  
\begin{equation}
\label{eta}
 \eta = \frac{z}{\sqrt{|b|} P |\Psi_H|} = \frac{ \frac{1-b}{2} \Psi_0 + \frac{b}{2} \alpha_Z}{\sqrt{|b|} P |\Psi_H|}
\end{equation}

The Bragg {\it uncorrected} angle verifies the Bragg law $\lambda=2 d_{hkl} \sin\theta_B$. 
The angle at $\eta=0$ corresponds to the Bragg angle corrected by refraction $\theta_{Bc}$: 
\begin{equation}
  \theta_{Bc} \approx \theta_B + \frac{1-b}{b} \frac{\Psi_0}{2 \sin(2 \theta_B)}
\end{equation}
Bragg angle and corrected Bragg angle are equal only for Laue symmetric case ($b=1$). The {\it Darwin width} is the angular 
interval $(\Delta \theta)_{D}$ that corresponds to a $2\eta$ width centered at $\eta=0$: 
\begin{equation}
  \label{eta}
  (\Delta \theta)_{D} = 2 \left| \frac{P |\Psi_H|}{\sqrt{|b|} \sin(2\theta_B)} \right|
\end{equation}
{\color{black} The Darwin width corresponds to the total reflection zone for a non-absorbing thick Bragg crystal, and to the FWHM 
(Full Width at Half Maximum) for the non-absorbing Laue crystal. It is often 
used as an indicator of the width of the diffraction profile. The Darwin width is exactly the FWHM for Laue non-absorbing crystals, 
but for the Bragg the case of thick-nonabsorbing crystals (known as Ewald solution) it is 
$FWHM=2\sqrt{3}/3 (\Delta \theta)_{D}\approx 1.155 (\Delta \theta)_{D}$. 
(or  $FWHM=3\sqrt{2}/4 (\Delta \theta)_{D}\approx 1.061 (\Delta \theta)_{D}$ for the small-absorbing infinitely-thick crystal, 
or Darwin solution) \cite{Zachariasen}. 
For a general crystal the FWHM can be calculated numerically from the simulated diffraction profile. }

Another important parameter is the extinction depth. The extinction in crystals is associated to the crystal 
thickness needed to diffract most of the beam, and it is associated to the primary extinction coefficient $\mu_{ext}$ or 
attenuation of the incident beam due to the diffraction. The extinction {\it depth} $\Lambda$ is the depth at which the 
{\it intensity} of the incident  wave is attenuated by a factor $1/e$:   

\begin{equation}
  \label{extinction}
  \Lambda = \frac{1}{\mu_{ext}} = \frac{\lambda |\gamma_0|}{2 \pi \sqrt{|b|} P |\Psi_H| \sqrt{1-\eta^2}}
\end{equation}

Its value is usually given at the center of the diffraction profile ($\eta=0$). In some cases, 
the extinction is given for {\it amplitude} instead of intensity, and the value is twice the one defined in 
Eq. (\ref{extinction}): $\Lambda_{ampl}=2\Lambda$. Moreover, one can also define the extinction {\it length} along the incident beam path 
(instead of {\it depth} along the crystal normal), thus $\Lambda_{length}=\Lambda /|\gamma_0|$. 
Usually extinction is associated to crystals in Bragg geometry, but for weakly absorbing crystals (as for Laue crystals) 
it is more appropriated to use the Pendell\"{o}sung depth $\Lambda_{pend}$ ($\Lambda_{pend}=2\pi \Lambda$). 

\subsection{The Multilamellar (ML) method}
\label{multilamellar}

The main idea behind this method is to decompose the crystal (in the direction of beam penetration) in several layers of a 
suitable thickness. Each layer behaves as a perfect crystal, thus the diffracted and transmitted beams are
calculated using the dynamical theory for plane crystals. The different layers are misaligned one with respect to the others 
in order to follow the cylindrical surface of the crystal plate. This model was first introduced by \cite{White} and further 
developed by, among others, \cite{Egert} and \cite{Boeuf}. It has been used for optimization of monochromators
for inelastic scattering and coronary angiography applications \cite{Suortti,Erola}, and for simulating crystal analyzers 
for fusion plasma diagnostics \cite{Caciuffo} . 

In the ML method, the bent crystal of thickness $T$ is decomposed into a series of perfect crystal lamellae of 
constant thickness $\Delta T$. The value of $\eta$ is then a function of the depth $t$ ($t=x_3$ axis) in 
the crystal from the entrance surface, or
\begin{equation}\label{eta_ml}
\eta(t) = \eta(0) + c A,
\end{equation}
where $\eta(0)$ is the $\eta$ value (Eq.~(\ref{eta})) at $t=0$ or entrance surface, 
$A$ is the crystal thickness in units of extinction depth for the amplitude given by~\cite{Zachariasen} (Eq. 3.140) 
\begin{equation}
\label{bigA}
A = \frac{T}{2 \Lambda} = 
    \frac{\pi P |\Psi_H|}{\lambda \sqrt{|\gamma_0 \gamma_h|}} T.
\end{equation}
The parameter $c$ is a reduced curvature, the ``deformation gradient''. It is a constant for uniform bending, 
and its expression comes directly from Eq. (\ref{eta_ml}) (see also \cite{ML03,Taupin1964}):
\begin{equation}\label{cparameter}
c = \frac{\mathrm{d} \eta}{\mathrm{d} A}, 
\end{equation}
and depends on the bending geometry and elasticity parameters of the crystal. Appendix \ref{appendix:elasticity} 
gives an introduction to the elastic anisotropy in crystals. An general expression of $c$
for a doubly curved Laue or Bragg crystal is obtained in Appendix~\ref{appendix:ml_taupin}:

\begin{eqnarray}\label{cparametergeneric}
c = \frac{-2 b \Lambda}{\sqrt{|b|} P |\Psi_H|} [ 
A_1(s_{21} \frac{M_1}{I} + s_{22} \frac{M_2}{I} ) + \nonumber \\
A_2(s_{31} \frac{M_1}{I} + s_{32} \frac{M_2}{I} ) + 
A_3(s_{41} \frac{M_1}{I} + s_{42} \frac{M_2}{I} ) ] 
\end{eqnarray}
where $M$ are the bending moments, $s_{ij}$ are components of the compliance tensor (see Appendix~\ref{appendix:elasticity}),
$I$ is the inertia moment of the crystal, and $A_i$ coefficients depending on the in and out beam directions:

\begin{eqnarray}\label{As} 
A_1 = (V^H_2)^2 V^0_3 - (V^0_2)^2 V^H_3 \nonumber \\
A_2 = V^H_3 V^0_3 (V^H_3-V^0_3) \\
A_3 = V^H_3 V^0_3 (V^H_2-V^0_2) \nonumber
\end{eqnarray}

\begin{figure}
\label{fig_ml}
\centering
\includegraphics[keepaspectratio,height=1.25in]{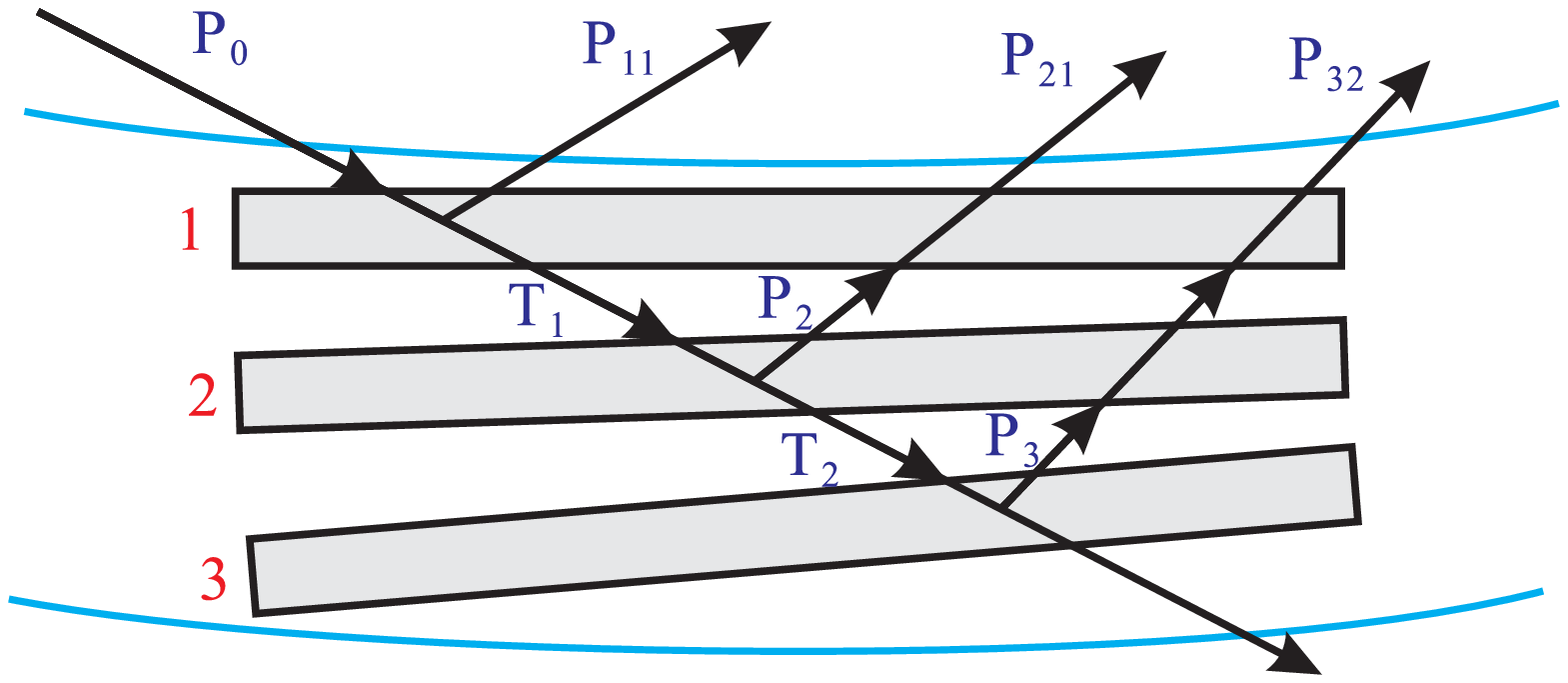}
\includegraphics[keepaspectratio,height=2.25in,width=2in]{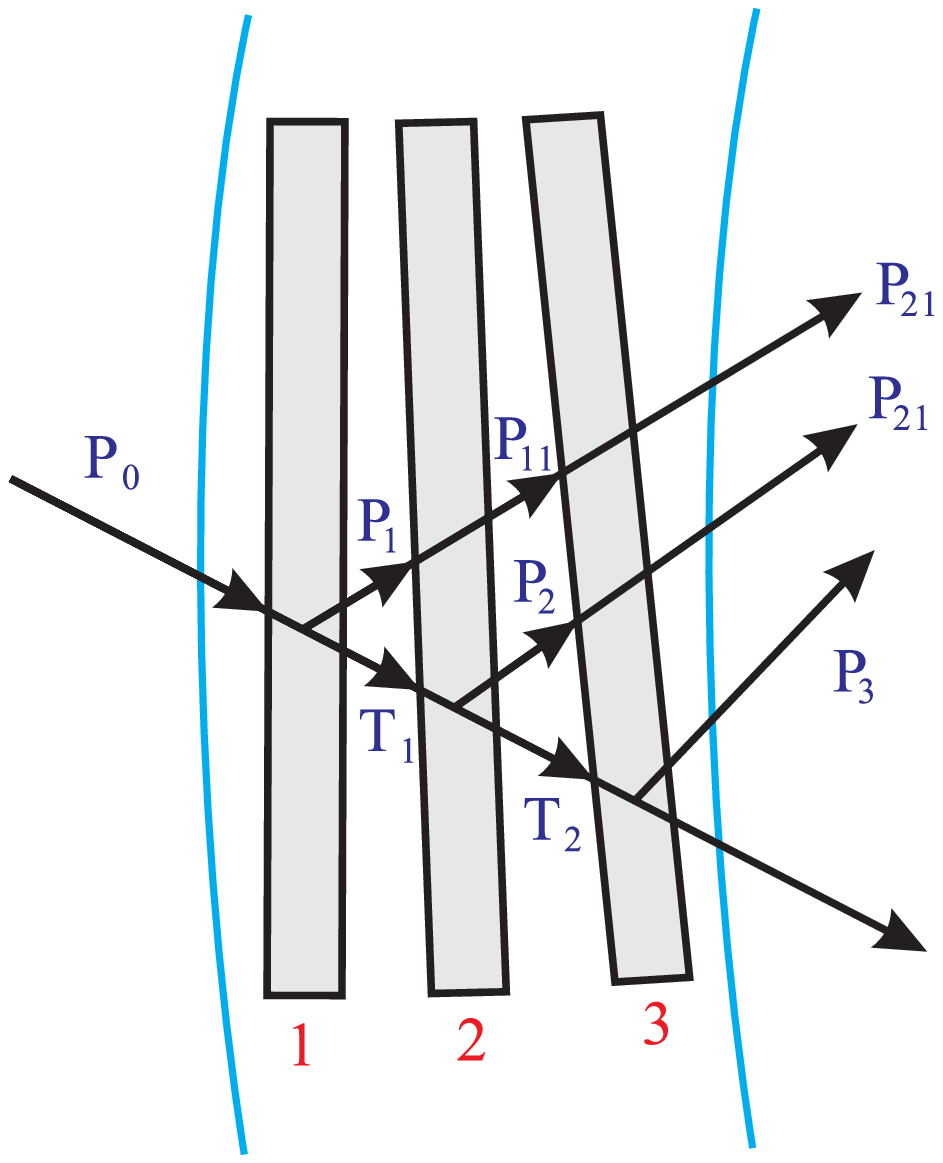}
\vspace{4mm}
\caption{Crystal division in the Multilamellar model for a Bragg and Laue crystal with the respective 
transmitted and diffracted beams for Bragg (top) and Laue (bottom) geometries.}
\end{figure}

Based on the adopted model (see Fig. \ref{fig_ml}), the Bragg planes in each lamella are tilted relative to 
the ones in its neighbor lamella by an angle $\Delta \eta = \pi /2$ for the Laue case and $\Delta \eta = 2$ 
for the Bragg case. Therefore, the reduced thickness for the lamella is $\Delta A = \Delta\eta/|c|$ (Eq. \ref{cparametergeneric}), 
thus giving $\Delta A = \pi/(2|c|)$ for Laue and $\Delta A = 2/|c|$ for Bragg case. The thickness of a lamella
is (Eq. \ref{bigA}) $\Delta T=2 \Lambda~\Delta A$ and the number of lamellae is $N=A/\Delta A=T/\Delta T$.

The total reflectivity of the given set of layers can be computed by writing the energy balance for the $n$ layers,
which leads for the Bragg case to:
\begin{equation}
\label{eq:ML5}
\mathcal{R} = \sum_{j=1}^{N} \left\{ r_j^{bragg} e^{ - \mu (j-1)S_H  }
\prod_{k=0}^{j-1} t_k^{bragg} \right\},
\end{equation}
and for the Laue case: 
\begin{equation}\label{eq13}
\mathcal{R} = \sum_{j=1}^{N} \left\{ r_j^{laue} e^{-\mu (n-j)S_H} \prod_{k=0}^{j-1} t_k^{laue} \right\},
\end{equation}
where $S_H = \Delta T/\gamma_h$ is the X-ray path of the diffracted beam inside a single lamella,
$\mu$ is the absorption coefficient of the crystal material, $r_i$ and $t_i$ are the  reflectivity and transmission 
for the $i$-th layer, respectively, that are computed using Eqs. (\ref{ZacBragg}) to (\ref{ZacTransLaue}), 
depending on the geometry (Bragg or Laue).

In the ML model it is assumed that the beam trajectory inside the crystal is a straight line. The crystal is perfect 
inside a given lamella, thus the crystal curvature cannot be large, otherwise it would originate local strains in 
the crystalline planes. The model is valid for crystals sufficiently thick to guarantee the existence of several 
lamellae. The model fails if the crystal thickness is of the order of or smaller than the lamella thickness. 
This may occur in Laue cases, where the crystal should be thin enough to guarantee a high transmission. 
The oscillations that may be found in Laue profiles calculated with the ML model are due to unphysical interferences 
between the crystal lamellae. A detailed description of the assumptions of this method is in~\cite{caciuffo_pr} 
and \cite{chantler} and a good validation with experiments is in~\cite{Erola}

\subsection{The Penning-Polder (PP) method}
\label{penningpolder}

The PP theory is another geometrical approach widely used because of its simplicity in generating the beam trajectory inside 
the crystals. In the original model \cite{PP01}, only the two ``normal wavefields" are accounted for. It is only valid when 
the disorientation of the crystal lattice planes over the Pendell\"{o}sung length (or extinction length) is much smaller than 
the intrinsic reflection width of the perfect crystal. However, by including the two ``created wavefields" from the 
interbranch scattering, the PP theory can be extended to the case of strongly distorted crystals \cite{PP03,Schulze}.

In the ray-optical theory of \cite{PP01} the X-ray beam in a distorted crystal is assumed to be a 
pseudo-plane Bloch wave ("wavefield ray") propagating parallel to the local Poynting vector. The crystal is supposed to 
be composed of {\color{black} parts of flat and undistorted crystals }where the dynamical theory for perfect crystals can be applied. 
The wavefield is preserved passing from one part  of the crystal to the next. Diffraction phenomena (the interference between two wavefields) 
are neglected. Thus, the  Pendell\"{o}sung fringes are not simulated with this model.

For crystals under constant strain gradient $\beta$ (see Appendix \ref{appendix:pp}) through the crystal thickness $T$, 
the ratio $\xi$ of the amplitudes of the diffracted and the transmitted waves can be obtained from Eq. 33 in \cite{PP01}
\begin{equation}\label{eq14}
2\eta = \xi_e - b/\xi_e = \xi_i - b/\xi_i + 2\beta T,
\end{equation}
where the subscript $e$ stands for the ``exit" surface, $i$ denotes the ``incident" surface.

The above equation generates two solutions for $\xi_i$ and another two for $\xi_e$:
\begin{gather}\label{eq14}
\xi_{e,1} = \eta + \sqrt{\eta^2-b} \nonumber \\
\xi_{e,2} = \eta - \sqrt{\eta^2-b} \nonumber  \\
\xi_{i,1} = \eta - \beta T + \sqrt{(\beta T - \eta)^2-b} \\
\xi_{i,2} = \eta - \beta T - \sqrt{(\beta T - \eta)^2-b} \nonumber
\end{gather}
these two solutions correspond to the splitting of the wavefields at the incident ($\xi_{i,j}, j=1,2$) and exit surfaces 
($\xi_{e,j}$). Most of the intensity goes along one mode for plane waves~\cite{PP03}. For each mode, the relationships between 
the intensity of the incident beam ($I_0$), the diffracted beam ($I_R$) and the transmitted beam ($I_T$) are
(Eq. 35 in \cite{PP01}):

\begin{gather}
\label{eq15}
\frac{I_{T,j}}{I_0} = \frac{b}{\xi_{i,j}^2+b}\frac{b}{\xi_{e,j}^2+b} \times \notag\\
\exp\left\{-\frac{\mu T}{\gamma_0} \left[ 1+\frac{b-1}{2\beta T}(\xi_{e,j} - \xi_{i,j}) + \frac{b}{\beta T} 
    \frac{\mathfrak{Im}\sqrt{\Psi_H \Psi_{\bar{H}}}}{\mathfrak{Im}\Psi_0}
\ln \frac{\xi_{e,j}}{\xi_{i,j}} \right] \right\}, \notag\\ 
\mathcal{R}_j = \frac{ I_{R,j} }{I_0} = \frac{\xi_{e,j}^2}{b} \frac{I_{T,j}}{I_0}.
\end{gather}

The total reflectivity $\mathcal{R}$ is then obtained by adding the intensities of the two $\xi$ solutions (one is
usually very small), and removing the intensity of the created wavefields from the interbranch: 
scattering \cite{PP03}, or 
\begin{equation}\label{eq16}
\mathcal{R} = \mathcal{R}_1  + \mathcal{R}_2  = \left( \frac{I_{R,1}}{I_o} + \frac{I_{R,2}}{I_o} \right) 
\left[1-\exp \left( -\frac{2 \pi \beta_c}{|\beta|} \right)\right],
\end{equation}
where $\beta_c = \pi / ( 2 \Lambda_{pend})$ is the critical strain gradient introduced by \cite{Authier1970},
with $\Lambda_{pend}$ the Pendell\"{o}sung period as defined in Sec. \ref{undistorted}. 
In case of overbending ($|\beta|>>\beta_c$), the $\mathcal{R}_j$ beams are attenuated and the intensity flows along
the transmitted beam direction:
\begin{equation}\label{eq16}
\mathcal{R}_j^{new} = \frac{I_{R,j}}{I_o} \exp \left( -\frac{2 \pi \beta_c}{|\beta|} \right).
\end{equation}

\section{Examples of application}

\subsection{Cylindrically bent crystals in Bragg geometry}

\begin{figure}
\label{fig_erola}
\begin{overpic}[width=3.5in] 
{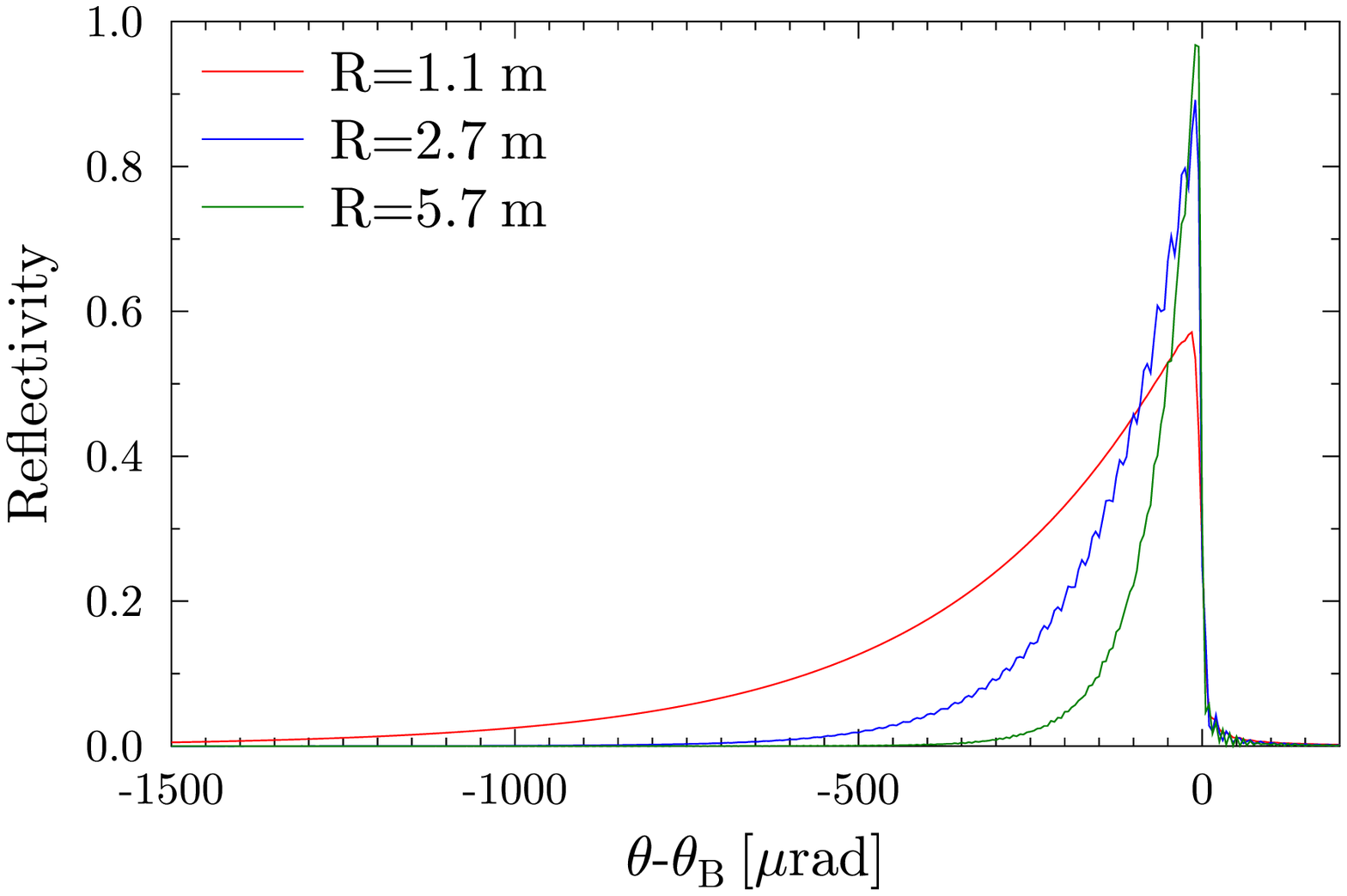}
\put(50,15){\includegraphics[width=1.5 in]{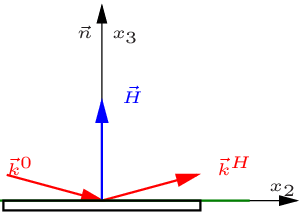}}
\end{overpic}
\caption{Calculated reflectivity curves for an isotropic symmetrical Si400 crystal reflection at photon 
energy $E=17479~eV$ ($\sigma$-polarized) for three curvature radii: $R=5.7~m$ (solid line), $R=2.7~m$ (dashed line) and $R=1.1~m$. 
The inset displays a sketch of the geometry. Results are in agreement with \cite{Erola}.} 
\end{figure}

In a first example we checked the output of our code against the experimental case described in Fig.~4 of \cite{Erola}, 
where it is shown the reflectivity for symmetrical Si400 reflection using Mo K$\alpha_1$ radiation 
($E=17479~eV$) for three bending radii: 1.1, 2.7 and 5.7~m. The bending is in the meridional
plane, and the crystal is considered isotropic (Poisson's ratio $\nu_{21}=-s_{21}/s_{22}=0.28$). 
The results of our simulations are in Fig.~\ref{fig_erola}. The results including crystal anisotropy are almost 
identical to the isotropic case. 
 
\subsection{Laue cylindrically bent crystal in meridional plane}

\begin{figure}
\label{fig_clemens_comparison_iso}
\begin{overpic}[width=3.5in] 
{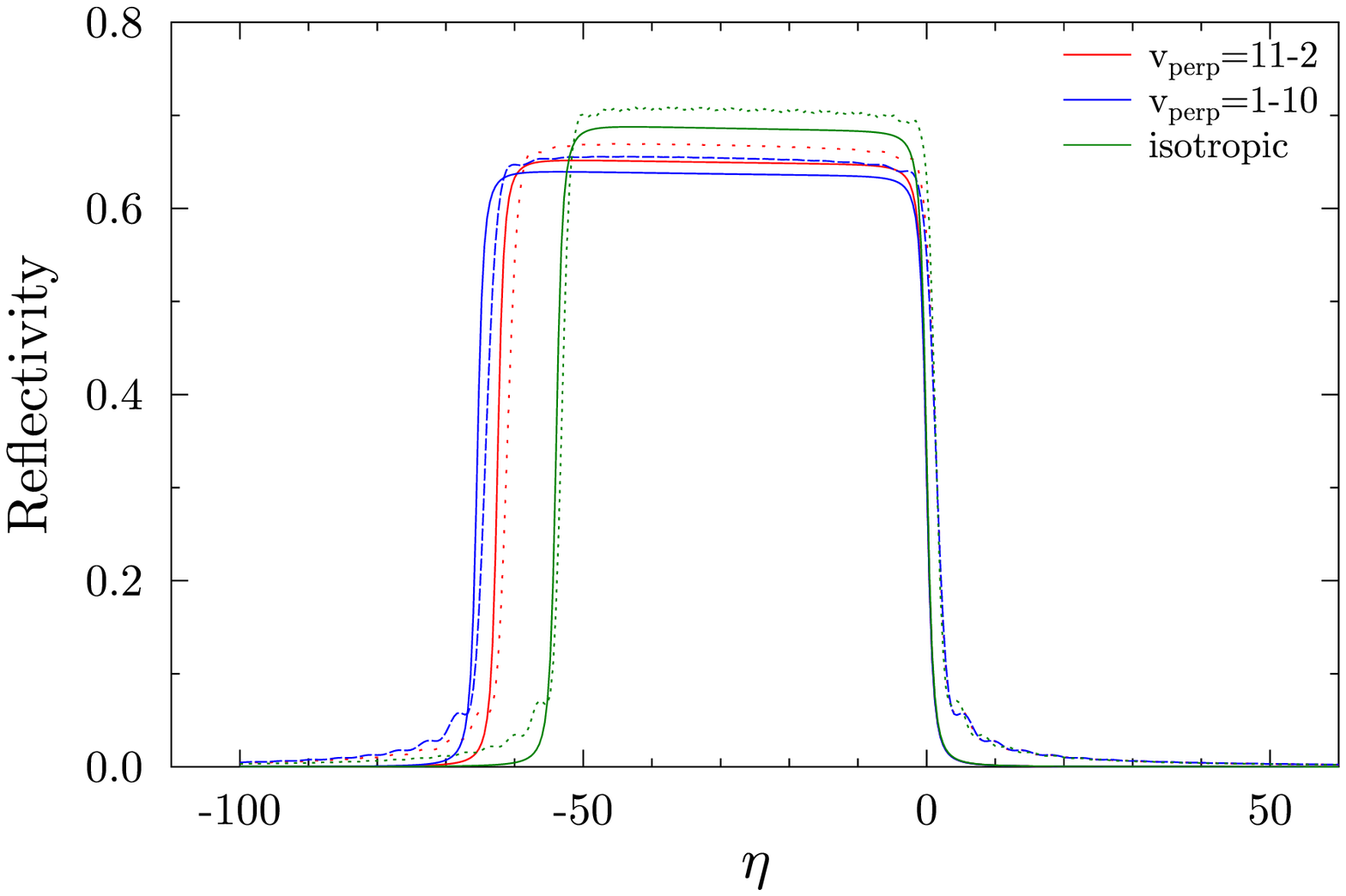}
\put(185,35){\includegraphics[width=1.0 in]{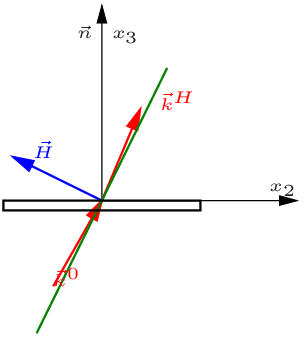}}
\end{overpic}
\caption{Diffraction profiles for a meridionally bent ($R$=325~cm), $T=700~\mu m$ thick silicon crystal at $E=33170~eV$. 
The calculations show the results for the asymmetric ($\alpha_X=296.2^{\circ}$) 111 reflection 
for the isotropic crystal (green lines, $\nu=0.274$) and two different crystal cuts:  
i) $\vec{v}_{\perp}= 1~1~\bar{2}$ (red lines) and ii) $\vec{v}_{\perp}= 1~\bar{1}~0$ (blue lines). 
The solid lines have been calculated using the Penning-Polder model and the dashed lines used the multilamellar model. 
In this case, 1$\eta=3.87\mu rad$. The inset displays a sketch of the geometry. 
}
\end{figure}

\begin{figure}
\label{fig_clemens_xsurface}
\caption{Diffraction profiles computed using the Penning-Polder model for a crystal with parameters like in 
Fig.~\ref{fig_clemens_comparison_iso} ($\vec{v}_{\perp}= 1~1~\bar{2}$) as a function of the bending radius.
}
\includegraphics[keepaspectratio,width=3.5in]{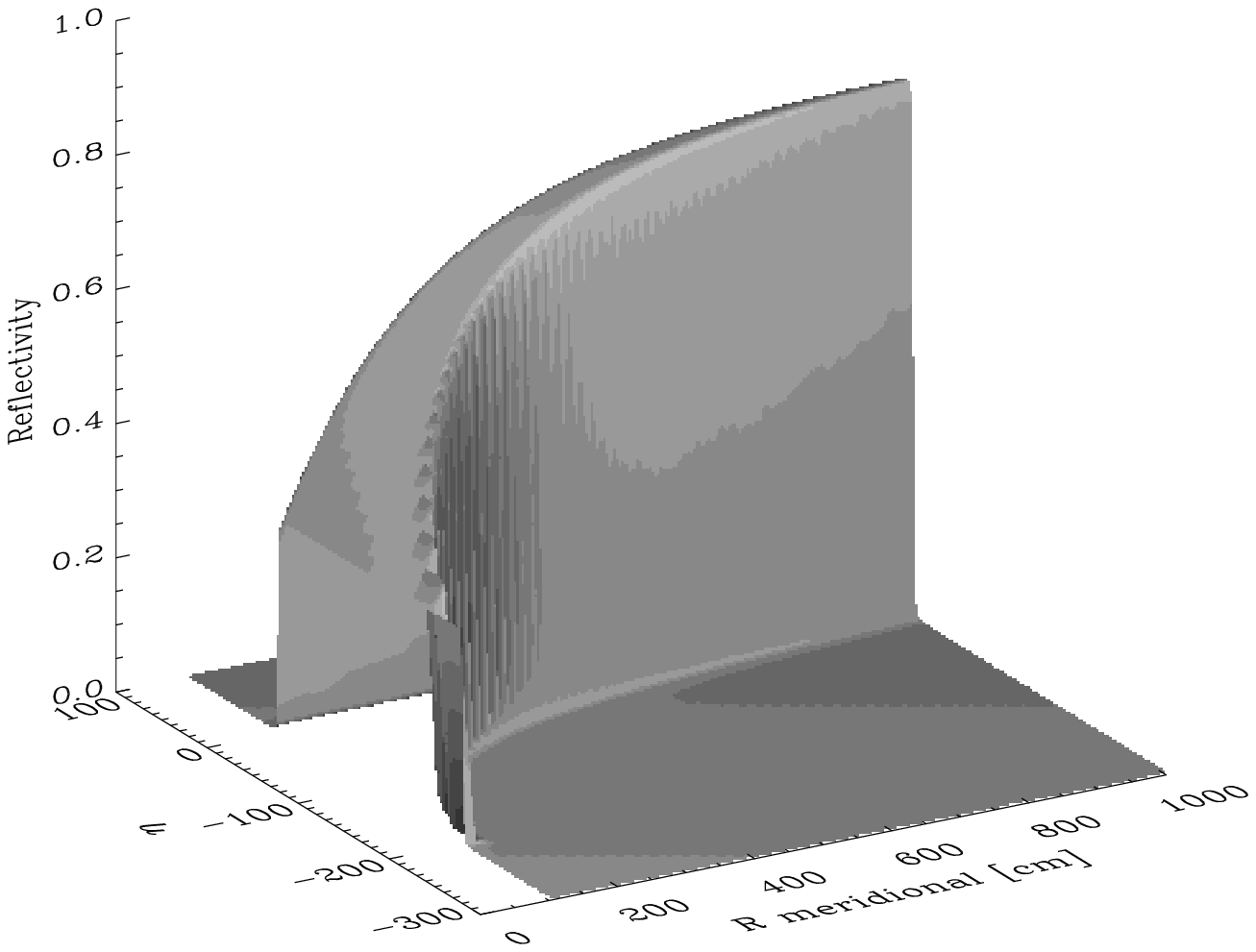}
\end{figure}

We calculate here the case described in \cite{Schulze} and treated extensively in \cite{SchulzeThesis}.
It consists of a Si111 crystal of thickness $T=700~\mu m$ diffracting at $E=33170~eV$, bent with
meridional Radius $R=325~cm$ (convex to the beam). The crystal directions for $\alpha_X=0$ are
$\vec{n}=1~1~1$, $\vec{v}_{along}=\bar{1}~1~0$, and $\vec{v}_{\perp}= 1~1~\bar{2}$. The 
asymmetrical cut is $\alpha_X=296.2^{\circ}$ ($\alpha=-116.2^{\circ}$, $\chi=-206.2^{\circ}$).
As discussed by Schulze, the anisotropy in the crystal is an important factor to be considered. 
Figure~\ref{fig_clemens_comparison_iso} illustrates this by comparing the calculated diffraction profiles for three 
cases: i) the isotropic crystal, ii) the original crystal configuration (with $\vec{v}_{\perp}= 1~1~\bar{2}$), and 
iii) another crystal cut ($\vec{v}_{\perp}=1~\bar{1}~0$).
It can be shown here that the curves generated by the two methods are in quite good agreement.
Fig.~\ref{fig_clemens_xsurface} shows the variation of the diffraction profile as a function of the bending radius.

\begin{figure}
\label{s_vs_alpha_play}
\caption{Variation of the elements of the compliance tensor that affect diffraction for 
meridional bending versus asymmetry angle $\alpha_X$ or $\chi$.
Continuous lines: elements affecting $c$ and $\beta$. Dotted line: element affecting anticlastic bending. 
}
\includegraphics[keepaspectratio,width=3.5in]{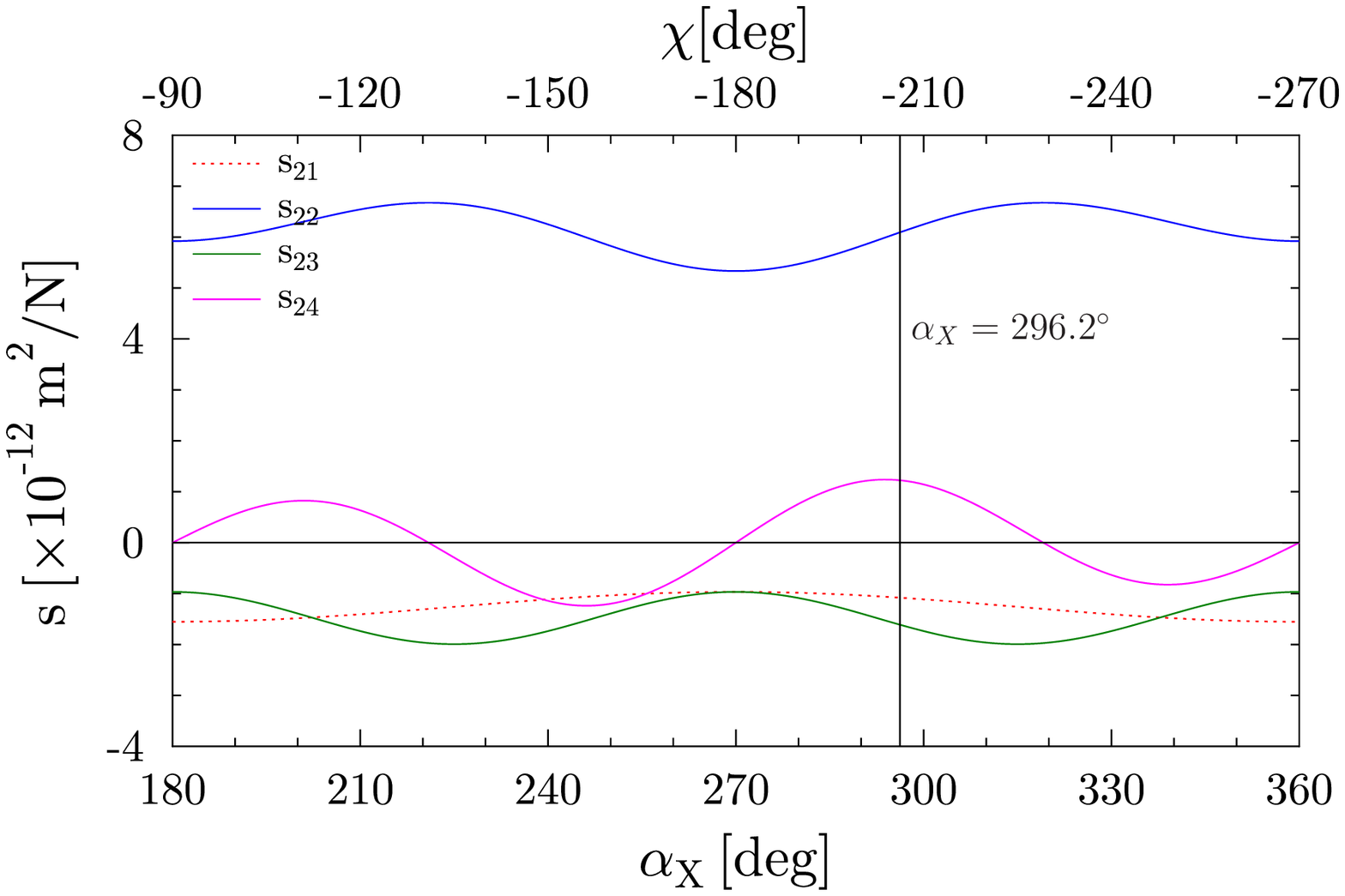}
\end{figure}

The compliance tensor for the chosen crystal cut and asymmetry angle is shown in Table~\ref{s_clemens}. The variation of 
the components of the compliance tensor versus asymmetry angle that affect the diffraction profile are in Fig.~\ref{s_vs_alpha_play}.

\begin{table}
\label{s_clemens}
\caption{Compliance tensor values ($\times 10^{-12}~m^2/N$) for an asymmetric ($\alpha_X=296.2^{\circ}$) silicon 111 crystal. 
In {\bf bold} the elements that affect the 
$\beta$ and $c$ parameters for pure meridional bending. 
The underlined elements  contribute to the anticlastic bending.
} 
\vspace{0.3cm}
\begin{tabular}{ccccccc}      
\hline
$j$ &  $s_{1j}$ & $s_{2j}$ & $s_{3j}$ & $s_{4j}$ & $s_{5j}$ & $s_{6j}$ \\

1&     5.920      &   \underline{-1.081}  &  -1.439  &  -0.465 &   -0.733  &   1.489 \\
2&    \underline{-1.081} &  {\bf \underline{6.092}}  &  {\bf -1.611}  & {\bf 1.225} &   -1.037  &  -0.871 \\
3&    -1.439 &  {\bf  -1.611}  &   6.450  &  -0.760 &    1.769  &  -0.618 \\ 
4&    -0.465 &  {\bf   1.225}  &  -0.760  &  14.715 &   -1.236  &  -2.074 \\
5&    -0.733 &   -1.037  &   1.769  &  -1.236 &   15.404  &  -0.930 \\
6&     1.489 &   -0.871  &  -0.618  &  -2.074 &   -0.930  &  16.836 \\
\hline     
\end{tabular}
\end{table}

\subsection{Optimization of a Laue cylindrically bent crystal in meridional plane}

\begin{figure}
\label{fig_veijo111}

\begin{overpic}[width=3.5in] 
{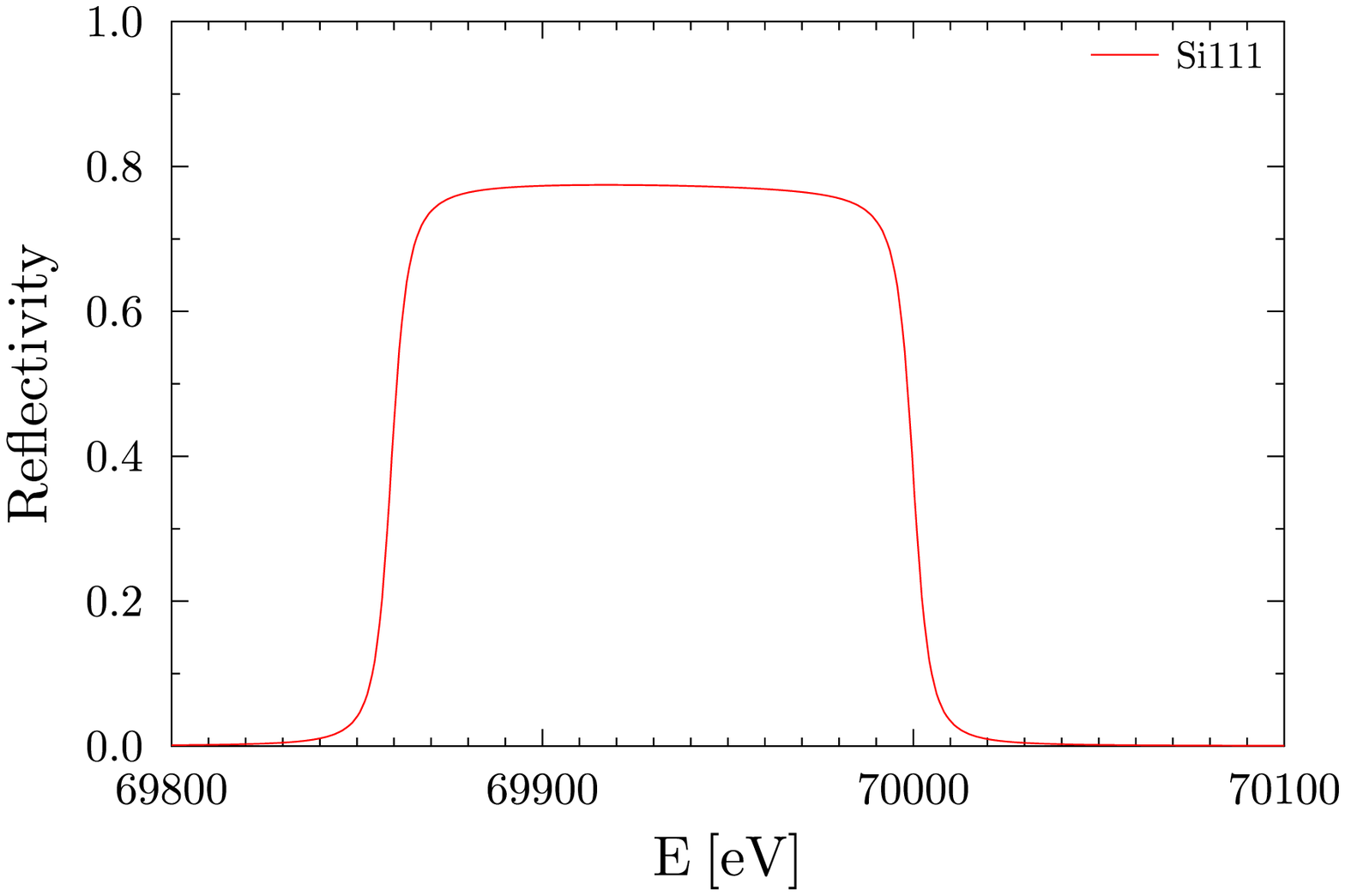}
\put(175,50){\includegraphics[width=1.25 in]{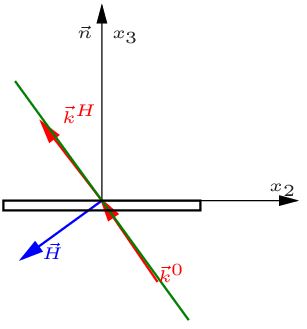}}
\end{overpic}
\caption{Diffraction profile for the Si111 crystal with parameters in Table~\ref{table_veijo}. }
\end{figure}

\begin{figure}
\label{fig_veijo113}
\begin{overpic}[width=3.5in] 
{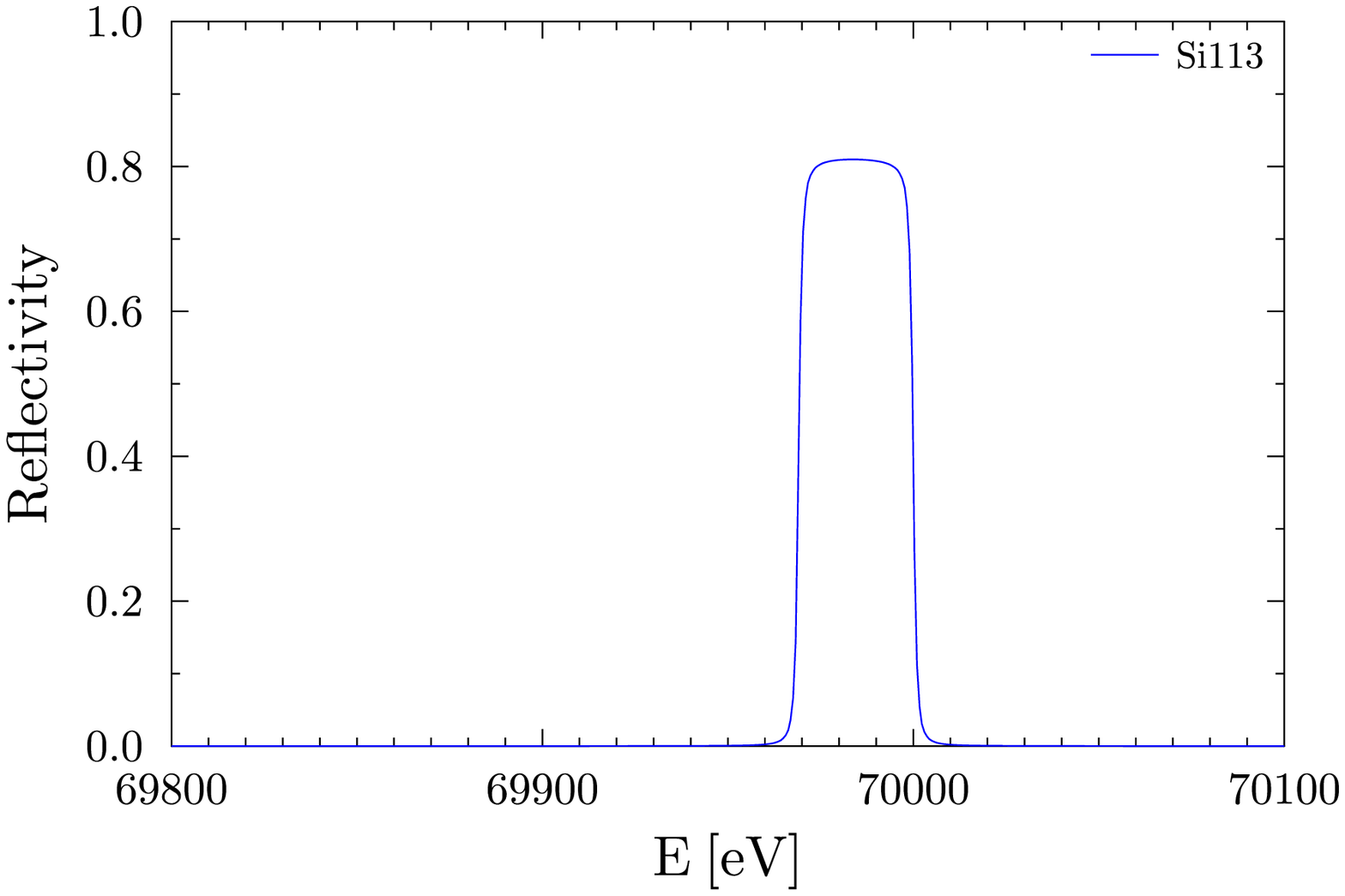}
\put(50,50){\includegraphics[width=1.25 in]{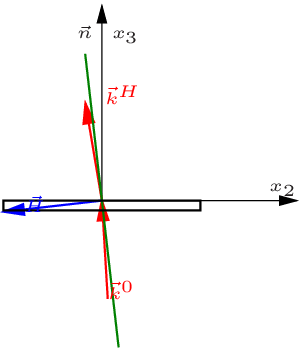}}
\end{overpic}
\caption{Diffraction profile for the Si113 crystal with parameters in Table~\ref{table_veijo}. }
\end{figure}

We calculate here the diffraction profile for Laue crystals to be used by the high energy X-ray monochromator 
proposed for the Upgrade ESRF beamline ID31 (UPBL02) \cite{UPBL02}. The monochromator holds two bent Laue crystals for monochromatizing 
the beam in the photon energy range 50-150~keV with variable energy resolution in the range $\Delta E/E\approx 10^{-4}-10^{-2}$.
controlled by the crystal bending. 
The diffraction plane is horizontal. We simulate a single silicon crystal at energy $E=70~keV$ using either the reflections 111 
(for low energy and low resolution) or 113 (for high resolution and high energy applications). 
The crystals parameters are in Table~\ref{table_veijo}. 
Fig.~\ref{fig_veijo111} shows the resulting diffraction profile for the 111 reflection and Fig.~\ref{fig_veijo113} 
considered crystal configuration.

\begin{table}
\label{table_veijo}
\caption{Inputs for a meridionally bent crystal.$R=10530cm/\gamma_0$ } 
\vspace{0.3cm}
\begin{tabular}{lcc}      
\hline
 Crystal    & Si~111 & Si~311   \\
 Photon energy    & \multicolumn{2}{c}{70 keV}  \\
 Crystal thickness  & \multicolumn{2}{c}{0.5 cm} \\
 Meridional Radius (convex to the beam) & 12759 cm & 10551 cm   \\
 Asymmetry $\alpha_X$     & $234^{\circ}$ &  $263.5^{\circ}$\\
 Asymmetry $\chi$     & $-144^{\circ}$ &  $-173.5^{\circ}$\\
 Crystal cut  (for $\alpha_X=0$) &  
\specialcell{$\vec{n}=1~1~1$ \\ $\vec{v}_{along}=\bar{1}~\bar{1}~2$ \\ $\vec{v}_{\perp}= \bar{1}~1~0$} & 
\specialcell{$\vec{n}=1~1~3$ \\ $\vec{v}_{along}=\bar{3}~\bar{3}~2$ \\ $\vec{v}_{\perp}= \bar{1}~1~0$}    \\
\hline
\end{tabular}
\end{table}

\begin{figure}
\label{fig_veijo_fig9}
\centering
\includegraphics[keepaspectratio,height=1.8in]{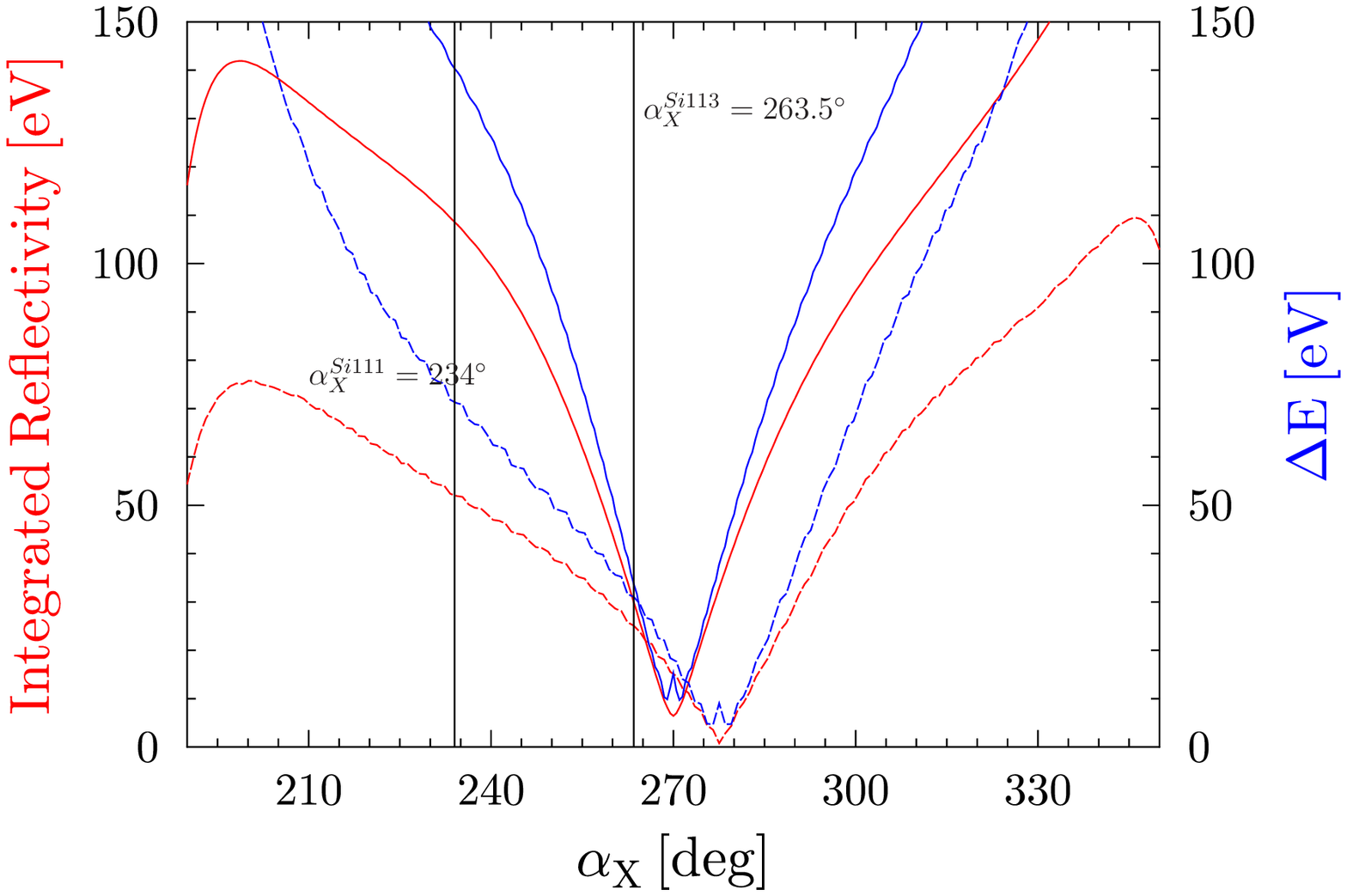}
~
\includegraphics[keepaspectratio,height=1.8in]{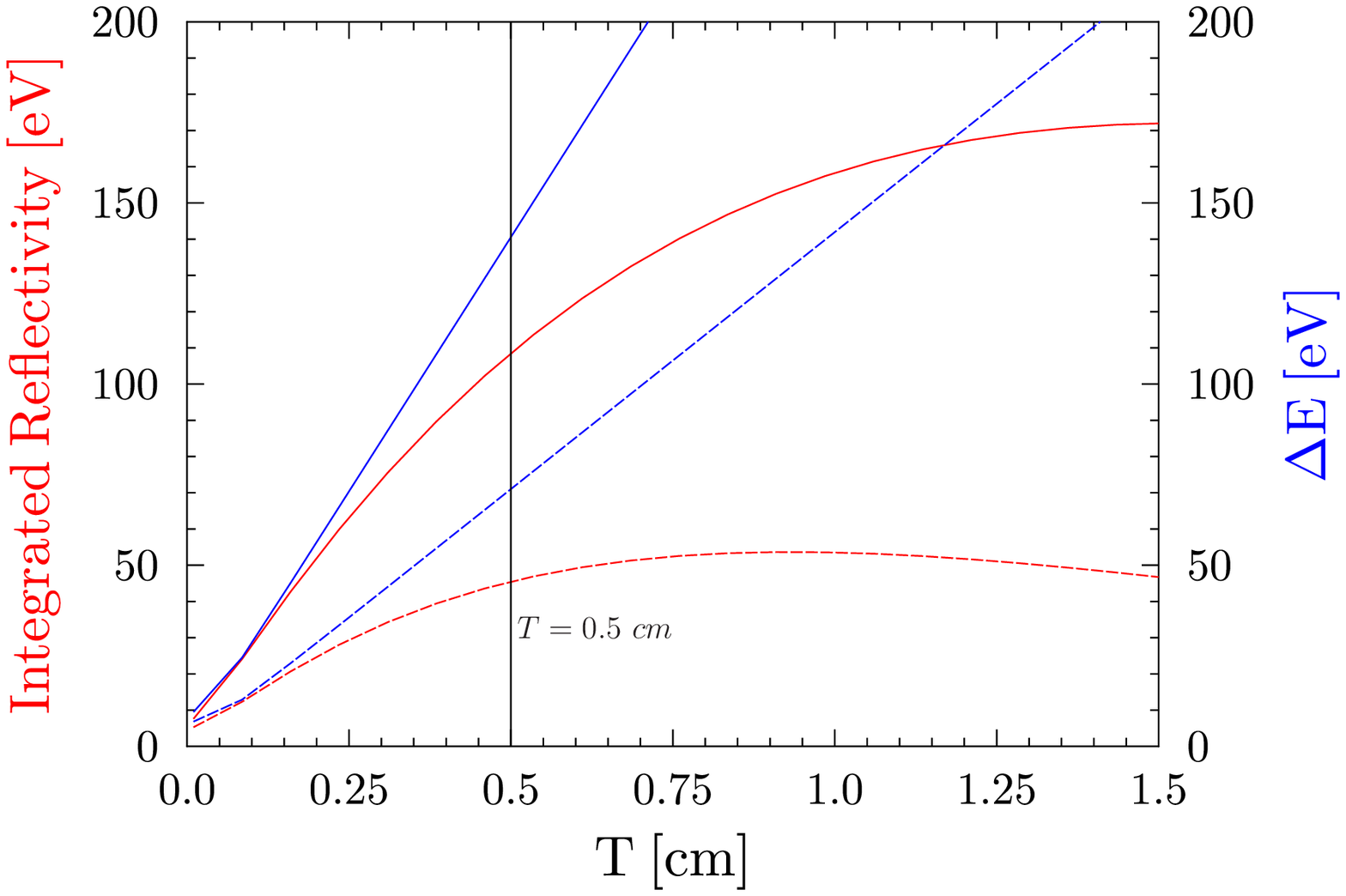}
\caption{Left: Dependency of integrated reflectivity (red) and $\Delta E$ (blue) for 
Si111 (solid) and Si113 (dotted) 
as a function of the asymmetry angle $\alpha_X$ (left). 
Right: Variation of integrated reflectivity (red) and $\Delta E$ (blue) for 
Si111 at 70 keV (solid) and 50 keV (dotted) as a function of crystal thickness (right).
}
\end{figure}

The crystal has been optimized in the following way: i) the crystal cut is chosen in such a way that 
$\vec{v}_{\perp}= \bar{1}~1~0$ so the compliance coefficient $s_{61}$ and $s_{62}$ 
are zero for any asymmetry angle. Therefore, the crystal is not twisted thus keeping $\vec{k}^0$, $\vec{k}^H$and $\vec{H}$
in the 23 plane. The crystal is cut with an asymmetry angle that on one side permits to use both Si111 and Si113 for low and high 
resolution applications, respectively; and on the other side is optimized for giving good integrated reflectivity.
Fig.~\ref{fig_veijo_fig9} shows the integrated reflectivity and energy bandwidth as a function of asymmetry angle and thickness, with indication 
of the selected working points. 

\begin{figure}
\label{fig_veijo_s}
\includegraphics[keepaspectratio,width=3.5in]{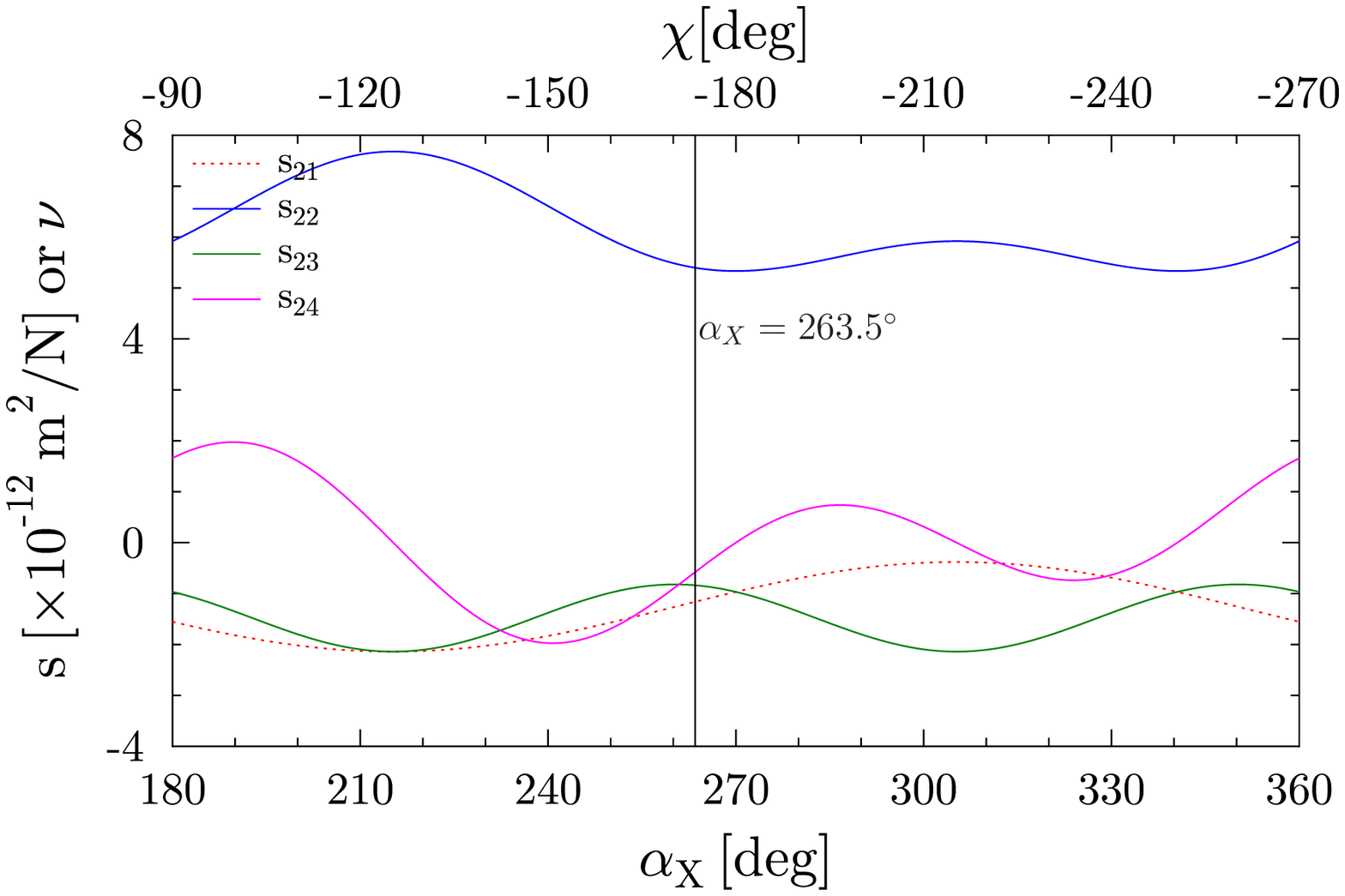}
\caption{Variation of the elements of the compliance tensor that affect diffraction for 
meridional bending versus asymmetry angle $\alpha_X$ or $\chi$ for Si111.
Continuous lines: elements affecting $c$ and $\beta$. Dotted line: element affecting anticlastic bending.  }
\end{figure}

The same physical crystal should be used for 111 and 113 reflections, and its compliance tensor
is shown in Table~\ref{s_veijo}. The variation as a function of the asymmetry (measured for Si111) of the components 
contributing to the diffraction is in Fig.~\ref{fig_veijo_s}.

\begin{table}
\label{s_veijo}
\caption{Compliance tensor values ($\times 10^{-12}~m^2/N$) for an asymmetric ($\alpha_X=234^{\circ}$) Si111 crystal
(or $\alpha_X=263.5^{\circ}$ Si113, corresponding to the same crystal cut). 
In {\bf bold} the elements that affect the 
$\beta$ and $c$ parameters for pure meridional bending. 
The underlined elements  contribute to the anticlastic bending.} 
\vspace{0.3cm}
\begin{tabular}{ccccccc}      
\hline
$j$ &  $s_{1j}$ & $s_{2j}$ & $s_{3j}$ & $s_{4j}$ & $s_{5j}$ & $s_{6j}$ \\
1&          5.920         &    \underline{-1.958}    &        -0.562  &       1.071  &   0.000 &    0.000  \\
2&    \underline{ -1.958} &  {\bf \underline{7.010}} &  {\bf  -1.651} & {\bf -1.810} &   0.000 &    0.000  \\
3&         -0.562         &  {\bf  -1.651}  &          5.613  &         0.739  &   0.000 &    0.000  \\
4&          1.071         &  {\bf  -1.810}  &          0.739  &        14.554  &   0.000 &    0.000  \\
5&          0.000         &         0.000   &          0.000  &         0.000  &  18.914 &    2.141  \\
6&          0.000         &         0.000   &          0.000  &         0.000  &   2.141 &   13.326  \\
\hline     
\end{tabular}
\end{table}

\subsection{Laue cylindrically bent crystal in sagittal plane}

\begin{figure}
\label{fig_xianbo_s}
\caption{Variation of Poisson's ratio versus asymmetry angle $\alpha_X$ (deg) }
\includegraphics[keepaspectratio,width=3.5in]{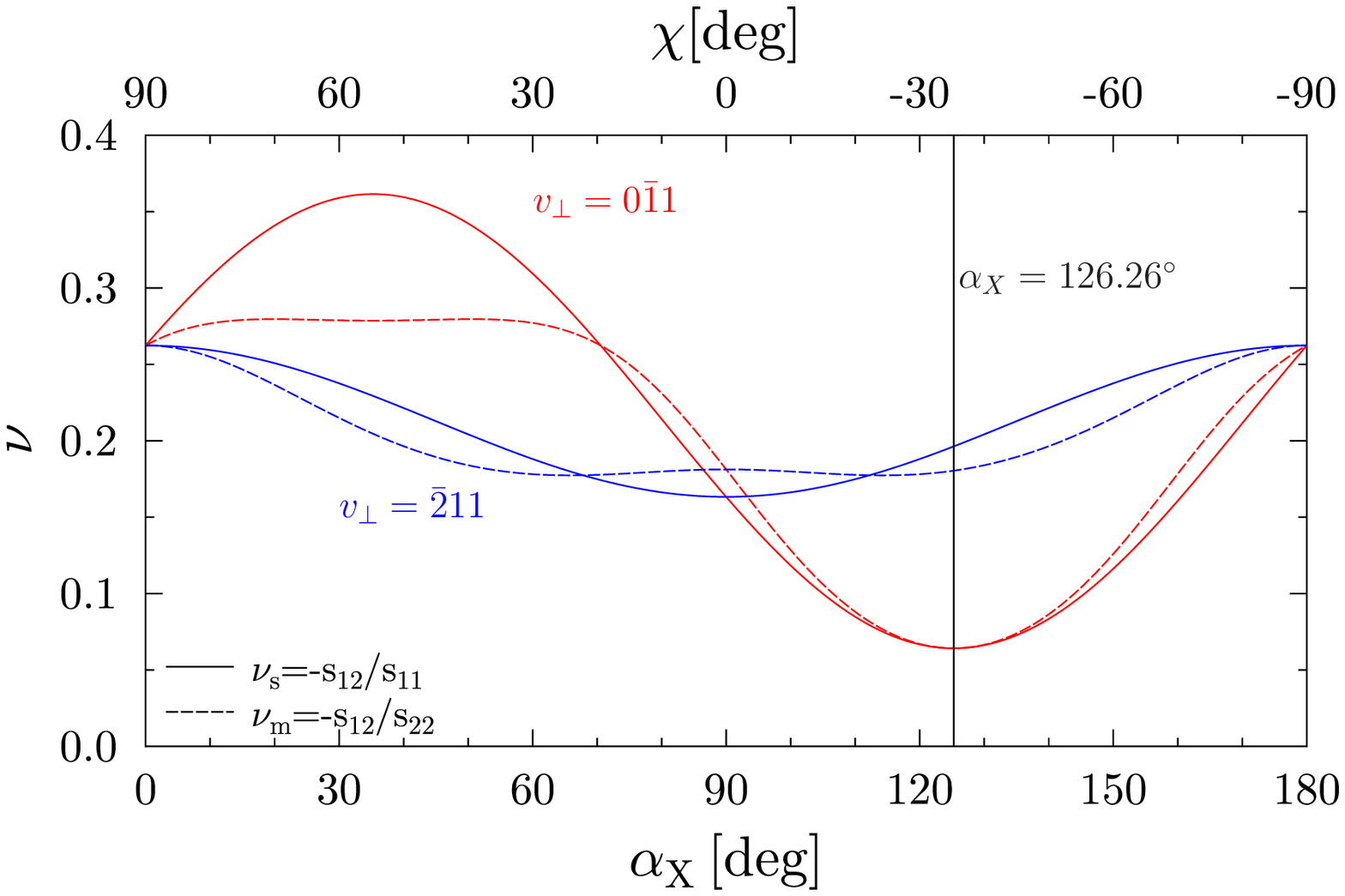}
\end{figure}

\begin{figure}
\label{fig_xianbo}
\begin{overpic}[width=3.5in] 
{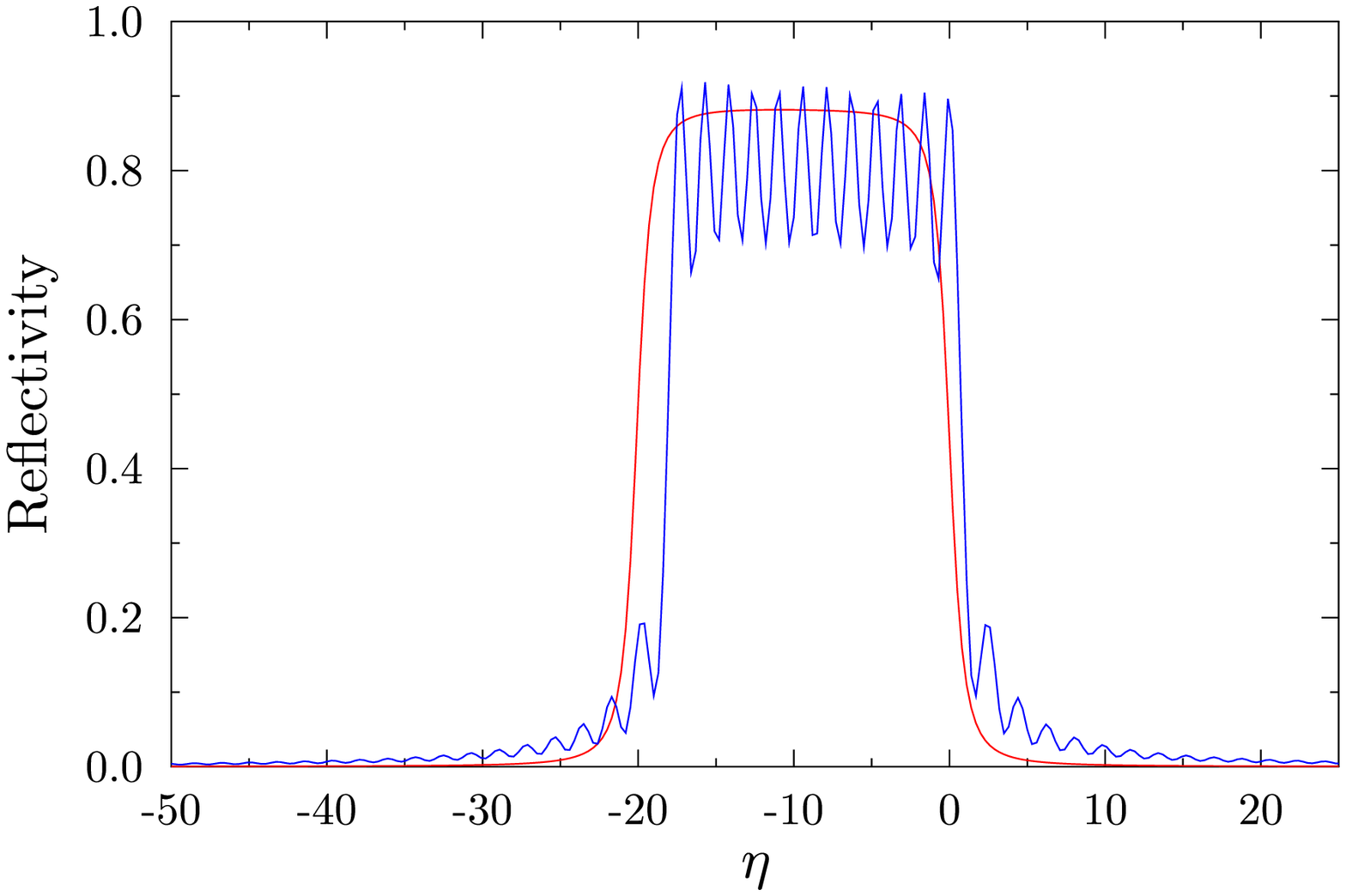}
\put(40,50){\includegraphics[width=1.25 in]{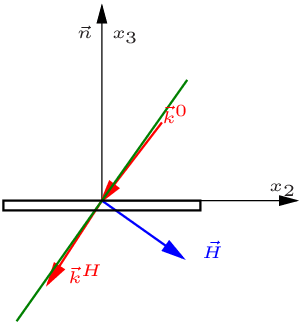}}
\end{overpic}
\caption{Diffraction profiles for a sagittaly bent Si$\bar{1}\bar{1}\bar{1}$ crystal (see text) calculated
using the Penning-Polder (red) and multilamellar (blue) models. In this case, $1\eta=2.57~\mu rad$.}
\end{figure}

In this section we analyze a Si$\bar{1}\bar{1}\bar{1}$ crystal 0.07~cm thick used to focus a 50 keV X-ray beam in sagittal direction \cite{xianbo_spie}.
The diffraction plane in the meridional direction is affected by the anticlastic curvature
originated by the sagittal radius, which is much smaller that the typical radii used in meridional focusing. 
One of the main roles of the crystal is to focus the beam in the sagittal direction, 
and to match the beam divergence in the meridional direction (Rowland condition). The optimized meridional radius $R_m$ is normally more 
than ten times larger than the sagittal radius $R_s$ required for focusing 
(e.g., in \cite{ShiJSR}). Therefore, in most cases the Poisson's ratio $\nu$ must be minimized 
($R_m=R_s/\nu$, with, $R_s=R_1$ the sagittal radius and $\nu=-s_{12}/s_{11}$). Fig.~\ref{fig_xianbo_s} 
shows the variation of the Poisson's ratio for different crystal cuts. From this graphic it is selected the 
crystal cut ($\vec{n}=\bar{1}~\bar{1}~\bar{1}$, $\vec{v}_{along}=\bar{2}~1~1$, and $\vec{v}_{perp}= 0~\bar{1}~1$ ) 
and asymmetry ($\alpha_X=125.26^{\circ}$ or $\chi=-35.26^{\circ}$). The compliance tensor is shown in Table~\ref{s_xianbo}.
Fig.~\ref{fig_xianbo} shows the diffraction profile. 

\begin{table}
\label{s_xianbo}
\caption{Compliance tensor values ($\times 10^{-12}~m^2/N$) for asymmetric ($\alpha_X=125.26^{\circ}$) 
Si$\bar{1}\bar{1}\bar{1}$ crystal. 
In {\bf bold} the elements that affect the 
$\beta$ and $c$ parameters for pure meridional bending. 
The underlined elements contribute to the anticlastic bending.} 
\vspace{0.3cm}
\begin{tabular}{ccccccc}      
\hline
$j$ &  $s_{1j}$ & $s_{2j}$ & $s_{3j}$ & $s_{4j}$ & $s_{5j}$ & $s_{6j}$ \\
1&   \underline { 5.920 } & {\bf \underline{-0.380} } & {\bf -2.140 } & {\bf   0.000 } &   0.000  &   0.000  \\
2&   {\bf \underline{ -0.380} } &   5.920  &  -2.140 &   -0.000  &   0.000  &   0.000  \\
3&   {\bf -2.140 } &  -2.140  &   7.680 &    0.000  &   0.000  &   0.000  \\
4&   {\bf  0.000 } &   0.000  &   0.000 &   12.600  &   0.000  &   0.000  \\
5&     0.000  &   0.000  &   0.000 &    0.000  &  12.600  &   0.000  \\
6&     0.000  &   0.000  &   0.000 &    0.000  &  -0.000  &  19.640  \\
\hline     
\end{tabular}
\end{table}

\section{Discussion and conclusions}

In the last section several diffraction profiles corresponding to Bragg and Laue geometries are considered. Numerical calculations
have provided the diffraction profiles. It is possible in some cases to obtain some physical parameters like the width of
the reflections, energy resolution and integrated intensities from simple analytical expressions obtained from the approximated 
methods applied. Some of these results are discussed in this paragraph. 

For Bragg curved crystals the multilamellar {\color{black} is the only method that can be applied among the ones described here.} 
The shape of the diffraction profile is approximately 
triangular in the case of ''thick`` crystals. The integrated reflectivity $R_\eta$ increases with curvature (inverse of radius). 
The integrated intensity (as a function of the $\eta$ variable) varies from  {\color{black} $R_\eta=\pi$ for the non-absorbing thick crystal 
($R_\eta=8/3$ for the Darwin solution)} to the kinematical limit $R_\eta^K=\pi^2 |\Psi_H|/(2 \lambda \mu)$ (mosaic crystals).
For instance, for the Si400 calculated in Fig.~\ref{fig_erola} $R_\eta^K=85.47$ and $R_\eta = 20.66, 36.44, 57.47$ for
radii $R=5.7, 2.7, 1.1~m$ (number of lamellae $1074, 2269, 5570$), respectively. 
In case of thin crystals, the diffraction profile does not decrease asymptotically to zero but decreases abruptly when at a 
given angle related to the crystal thickness. It produces a trapezoidal-shaped profile for Bragg crystals, and an almost 
rectangular shape for Laue crystals (as seen in all Laue examples discussed in last section). 
The width of the reflection is (from Eq.~\ref{cparametergeneric}) $\Delta \eta = c A=c T \Lambda /2$. The energy bandwidth 
is calculated using the derivative of the Bragg law and Eq.~\ref{eta}:  
\begin{equation}
 \label{DeltaE_ML}
 \frac{\Delta\lambda}{\lambda}=-\frac{\Delta E}{E}= \cot \theta_B \Delta \theta = \frac{\pi P^2 |\Psi_H|^2}{2 \lambda |\gamma_0| \sin^2\theta_B}c T
\end{equation}

Results of values given by Eq.~\ref{DeltaE_ML} to the Laue crystals discussed in the last section are shown in Table~\ref{table_discussion}.

For Laue crystals the PP method gives a diffraction profile width $\Delta \eta \approx \beta T$ in case of low-absorbing crystals 
curved enough to produce a diffraction width much larger than the {\color{black} perfect (undistorted) crystal}
($\Delta \eta \gg 2$). Using Eq.~\ref{beta_xianbo} it gives a bandwidth of:

\begin{equation}
 \label{DeltaE_XIANBO}
 \frac{\Delta\lambda}{\lambda}=-\frac{\Delta E}{E}=  \frac{G T}{2 k^0 \gamma_0 \sin^2 \theta_B}
\end{equation}

\begin{table}
\label{table_discussion}
\caption{Comparison of energy bandwidth for different Laue crystal reflections analyzed in previous section.}
\begin{tabular}{lcccc}      
                                    &  Fig.~\ref{fig_clemens_comparison_iso} &  Fig.~\ref{fig_veijo111}  &  Fig.~\ref{fig_veijo113} & Fig.~\ref{fig_xianbo} \\
\hline
$\Delta E$ FWHM from ML profile   &   142.9  &   138.0    &     30.6  &     64.8     \\
$\Delta E$ Eq. \ref{DeltaE_ML}    &   142.5  &   141.6    &     31.3  &     68.8     \\
\hline
$\Delta E$ FWHM from PP profile   &   143.5  &   140.6    &     30.9  &     69.0     \\
$\Delta E$ Eq. \ref{DeltaE_XIANBO}    &   143.5  &   141.3    &     31.1  &     68.8    \\
\end{tabular}
\end{table}

The numerical values for energy bandwidth given by the analytical formulas agree very well with the calculated ones. 
In fact, both models agree if $\beta T = c A$ which implies that 
\begin{equation}
\frac{d \alpha_Z}{dt} = \frac{1}{\gamma_0 k^0} \frac{\partial^2 (\vec{H}.\vec{u})}{\partial V_0 \partial V_H}
\end{equation}
Therefore, both models are consistent in giving the same $\Delta E$, they give approximated peak reflectivities
(checked numerically for the treated cases), but will produce different ray trajectories (not discussed here).

\begin{figure}
\label{fig_sphere}
\caption{Comparison of the reflectivity curve for the crystal in Fig. \ref{fig_xianbo} (pure sagittal bending,
$R_s=-125~cm$ solid line), with the pure meridional bending ($R_m=-125~cm$, dotted line), and spherical 
bending ($R_s=R_m=-125~cm$ dashed line). In red, the diffraction profile produced by a crystal bent in the 
meridional direction with $R_{anticlastic}=1947.37~cm$.
}
\includegraphics[keepaspectratio,width=3.5in]{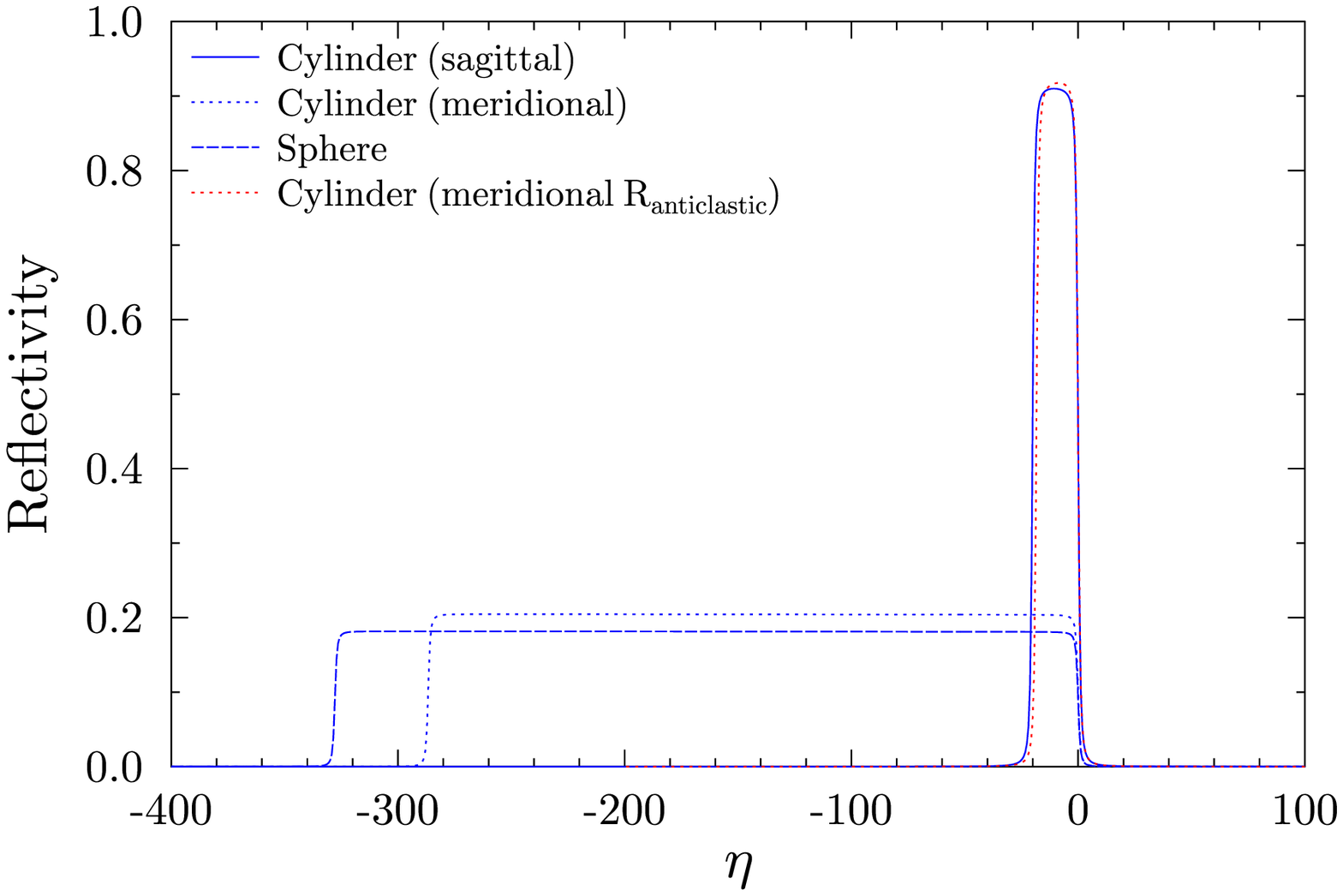}
\end{figure}

We compare now the diffraction profiles produced by the same crystal curved along the sagittal direction 
with the same crystal curved with the same radius $R=$-125~cm along the meridional direction, 
or in both direction (spherical). The results are shown in Fig.~\ref{fig_sphere} showing remarkable differences. 
The broadening of the diffraction profile is produced by the curvature in the diffraction plane, 
therefore the broadening is very important for pure meridional or spherical bending. 
For sagittal bending, the broadening is much smaller, because the curvature in the meridional plane is due to the larger 
anticlastic radius $R_{anticlastic}=-R_m/ \nu_s$, where $R_m$ is the meridional bending and 
$\nu_s=-s_{12}/s_{11}=0.064$ is the Poisson's ratio in sagittal direction. In fact, the diffraction profile calculation 
for sagittally bent crystal of $R=-125~cm$ is very well approximated by a meridional bending with 
$R_{anticlastic}=1947.37~cm$ (see Fig.~\ref{fig_sphere}).

{\color{black}
This paper concerns distorted crystals in which the deformation is created by bending the crystal, usually elastic bending. 
In most cases the crystal curvature is created with the aim of focusing or collimating the X-ray beam, usually in the 
meridional plane, but sometimes in the sagittal direction. We discussed the general case of curvature in both directions, which is of interest 
for spherical bending (equal two-moments), toroidal (different two-moments), or when cylindrical bending is requested (one-moment bending). 
In this last case, however, the elastic properties of the materials induce the anticlastic curvature in the perpendicular direction. 
Another important source of deformation for crystals in synchrotron beamlines is the thermal load. Although a full analysis of the 
crystal reflectivity for heat load in crystals is out of the scope of this paper, the same approximated methods to calculate
the diffraction profiles can be used. In fact, the original work of \cite{Penning} also gives the value of $\beta$ for a 
crystal deformed by a uniform temperature gradient. Also, our code can be used to estimate if the thermal deformation affects
the diffraction properties of the crystal by calculating the diffraction profiles for a curvature radius $R$ due to the 
thermal bending of the crystal \cite{Kalus1973}:

\begin{equation}
R = \frac{T}{\Delta_T \alpha_S}
\end{equation}

where $T$ is the crystal thickness, $\alpha_S$ is the thermal expansion coefficient of the crystal and 
$\Delta_T$ is the temperature difference between the two faces of the crystal that is approximately proportional to the 
absorber power  $\Delta_T=P T/\kappa$, with $\kappa$ the diffusion coefficient. This only gives a first estimation 
of the possible alteration of the diffraction profile. Other effects must be considered, like the thermal expansion of the crystal 
unit cell, and the changes in the rocking curves because of the reflection in a non-planar crystal surface \cite{ZhangMonochromators}.  
}

In conclusion, we summarized the formulation of two approximated methods for calculating diffraction profiles:
the multilamellar applied for both Bragg and Laue bent crystals, and the Penning-Polder only applicable to Laue crystals. 
We obtained the general expression of the constant strain parameter ($\beta$ for PP and $c$ for ML) including crystal anisotropy
for any asymmetric crystal cut. The general expressions obtained reduce to the formulations of particular cases from literature. 
Last, some approximated expressions for obtaining the angular and energy bandwidths of these crystals are given. 

{\color{black} The equations are implemented in the XOP package and are validated by studying some examples analyzed in literature. 
The XOP user has access to these calculations using the usual {\tt XCRYSTAL} application. The input files for {\tt XCRYSTAL} with the
input parameters of the cases studied in this paper are available in the {\tt examples} directory of the XOP distribution. 
The code is also provided in open source under the GPL license at {\tt https://github.com/srio/CRYSTAL}. }

\newpage

\appendix

\numberwithin{equation}{section}

\section{Geometry, scattering vectors and angles}
\label{preliminaries}

We define an orthonormal reference frame intrinsic to the crystal with origin at the crystal center (usually 
where the central ray intercept the crystal) and three vectors $(\vec{n},\vec{v}_{along},\vec{v}_{\perp})$, 
being $\vec{n}$ a unity vector normal to the crystal surface pointing outside the crystal bulk, 
and $\vec{v}_{along}$ and $\vec{v}_{\perp}$ 
two orthonormal vectors in the plane of the crystal surface, usually chosen to be inside and perpendicular
to the diffraction plane, respectively.  
%
A crystal cut is defined by the asymmetry angle, either $\alpha$  or $\chi$. 
Usually $\alpha$ is used in Bragg geometry and $\chi$ in Laue geometry, but if both values are 
well-defined they can be used indistinctly in both geometries.

\begin{figure}
\label{figlauebragg}
\centering
\includegraphics[keepaspectratio,height=2.15in ,width=8in]{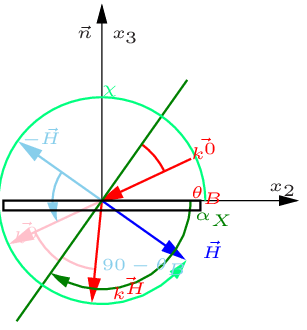}
~~~~~~~~~~~~~~
\includegraphics[keepaspectratio,height=2.15in,width=6in]{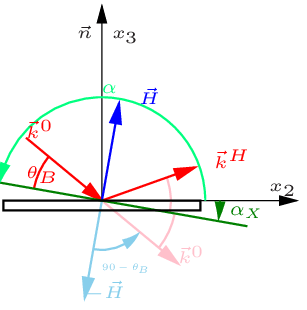}
\caption{Schematic view of diffraction using asymmetric Si111 crystals at $E=4~keV,~\theta_{B}=29.62^{\circ}$. 
Left: Laue geometry: $\alpha_X=125.26^{\circ}, \chi=-35.26^{\circ},~\alpha=54.74^{\circ}$.
Right: Bragg geometry: $\alpha_X=10^{\circ}, \chi=80^{\circ},~\alpha=170^{\circ}$.  
The plotted incident direction corresponds to the Bragg position $\vec{V}^0_B$, so here $\vec{k^0}=\vec{k^0_B}$.
}
\end{figure}

For the formulas used in this text, implemented in the XOP code, we have used the following definitions and
conventions: 
\begin{itemize}
 \item For simplicity, the three reference vectors previously defined match a simple laboratory 
reference system chosen $\vec{n}=(0,0,1)$, $\vec{v}_{along}=(0,1,0)$ and 
$\vec{v}_{\perp}=(1,0,0)$. The normal of the reflecting crystal surface $\vec{n}$ is pointing outside from the crystal. 
A generic point in the crystal has as coordinates $\vec{r}=(x_1,x_2,x_3)$.
 \item We define $\chi$ as the angle from $x_2$ axis to $\vec{H}$ (in mathematical sense, positive if counterclockwise (ccw)).
The angle $\alpha=\chi+90^{\circ}$ goes from the crystal surface to the the Bragg planes $hkl$\
(measured from $x_2$ axis to the reflecting surface, thus for symmetric Bragg case $\alpha=180^{\circ}$). 
The asymmetry angle $\alpha_X$ defined in XOP (from crystal surface to crystal planes, positive if clockwise (cw)) holds 
$\alpha_X+\alpha=180^{\circ}$.  See Fig. \ref{figlauebragg}.
 \item The reciprocal lattice vector $\vec{H}$ can be obtained by rotating the normal to the surface $\vec{n}$ 
an angle $\theta=\chi-90^{\circ}=-\alpha_X$ around
the vector $\vec{v}_{\perp}$: $\vec{H}=ROT_{\vec{v}_{\perp}} (\vec{n},\theta )$.
The $ROT$ operator is implemented via the Rodrigues 
formula that rotates a vector $\vec{V}$ an angle $\theta$ around an axis $\vec{a}$ (normalized, 
$\theta $ is positive in the screw (cw) sense when looking in the direction of $\vec{a}$) to obtain $\vec{V_{rot}}$:
\begin{eqnarray}
  \label{rodrigues}
  \vec{V}_{rot} = ROT_{\vec{a}}(\vec{V},\theta)=\vec{V} \cos\theta + \nonumber \\
  (\vec{a} \times \vec{V}) \sin\theta + 
  \vec{a}~(\vec{a} \cdot \vec{V})(1-\cos\theta).
\end{eqnarray}

 \item
The (unsigned) Bragg angle verifies $\lambda=2 d_{hkl} \sin \theta_B$, with $\lambda$ the photon wavelength in vacuum. 
The direction of an incident beam that fulfils Bragg's law makes an angle $\theta_{B}$ with respect to the Bragg planes, 
therefore its direction is:  
$\vec{V}^0_{B}=ROT_{  \vec{v}_{\perp}  }(-\vec{H},90^{\circ}-\theta_{B})$.
\end{itemize}

The diffraction profile for a given {\it hkl} reflection is obtained by scanning the direction $\vec{V}^0$ 
of a monochromatic collimated
incident beam (a plane wave) in the vicinity of $\vec{V}^0_{B}$ in the diffraction plane. If both vectors are separated by an angle
$\Delta\theta=|\theta|-|\theta_B|$ we can set $\vec{V}^0=ROT_{\vec{v}_{\perp}} (\vec{V}^0_{B},-\Delta\theta)$.

The vector expressions can be expressed in angles, but care must be taken with the definition and sign of angles. 

The incident angle (measured ccw from $x_2$ to $\vec{V}^0$) for the ray fulfilling the Bragg law (in both
Bragg and Laue geometries) is $\theta_{0}=180^{\circ}+\alpha-\theta_B$, and the reflected angle 
(measured ccw from $x_2$ to $\vec{V}^H$) is $\theta_H=180^{\circ}+\alpha-\theta_B$. 
In Laue geometry, for a given crystal cut, we may send the beam to match the Bragg angle 
below or onto the Bragg planes (which strictly speaking correspond to using $\pm\vec{H}$), and reversing time.
From these four cases only two are independent. 

The directions (as as function of $\alpha$, $\alpha_X=180^{\circ}-\alpha$, and $\chi=\alpha-90^{\circ}$),
for the directions fulfilling the Bragg law (Eq. \ref{eq:laue}) are: 

\begin{gather}
\label{vector2angles}
\vec{V}^0_B=(0,\cos\theta_0,\sin\theta_0)= \nonumber \\
  (0,-\cos(\theta_{B}-\alpha),\sin(\theta_{B}-\alpha))=  \nonumber \\
  (0,\cos(\theta_{B}+\alpha_X),-\sin(\theta_{B}+\alpha_X))= \nonumber \\
  (0,-\sin(\theta_{B}-\chi),-\cos(\theta_{B}-\chi))              \nonumber \\ \nonumber \\
\vec{V}^H_B=(0,\cos\theta_H,\sin\theta_H)= \nonumber \\
  (0,-\cos(\theta_{B}+\alpha),-\sin(\theta_{B}+\alpha))=  \\
  (0,\cos(\theta_{B}-\alpha_X),\sin(\theta_{B}-\alpha_X))= \nonumber \\
  (0, \sin(\theta_{B}+\chi),-\cos(\theta_{B}+\chi))               \nonumber \\ \nonumber \\
\vec{n}^H = (0,\sin\alpha,-\cos\alpha)= \nonumber \\
            (0,\sin\alpha_X,\cos\alpha_X) =  \nonumber \\
            (0,\cos\chi,\sin\chi) \nonumber
\end{gather}

\section{Elastically bent crystals}
\label{appendix:elasticity}

\subsection{Bending an anisotropic plate}

The generalized Hooke's law makes a ``linear'' relation between the the $strain$ tensor ($S'_{ij}$, $i,j=1,2,3$, characterizing 
the deformation, adimensional), and the $stress$ tensor ($\sigma'_{kl}$, $k,l=1,2,3$, generalized forces with dimension of 
force times $L^{-2}$)  (Eq. 1.3.2 in \cite{hearmon}): 
\begin{equation}
\label{hooke1}
S'_{ij} = s'_{ijkl} \sigma'_{kl}
\end{equation}
where $s'_{ijkl}$ is the $elastic~compliance$ tensor. The {\it stiffness} tensor is the inverse of the compliance tensor, but here
we will center the discussion on the compliance tensor only. Each index can take three values (1,2,3, corresponding to the three 
directions in 3D space) and the repeated indices are summed. From symmetry considerations, not all 81 elements of the compliance 
are independent, but they are reduced to 36. Moreover, thermodynamical considerations reduce the number of independent 
compliance elements to a maximum of 21 in the most general case. These facts make possible to reduce 
Eq. (\ref{hooke1}) to a simpler form  (Eq. 1.3.6 in \cite{hearmon}): 
\begin{equation}
\label{hooke2}
S_{q} = s_{qr} \sigma_{r}~~(q,r=1,...,6),
\end{equation}
with now $s_{qr}$ a 36 elements symmetric matrix. The new strain 6-dim vector as a function of the old strain tensor is 
 (Eq. 1.2.8 in \cite{hearmon}): 
\begin{eqnarray}
S_1=S'_{11},~S_2=S'_{22},~S_3=S'_{33}, \nonumber \\
~S_4=2 S'_{23},~S_5=2 S'_{13},~S_6=2 S'_{12},
\end{eqnarray}
(the same convention applies for indices from $s'$ to $s$, e.g., $s_{36}=s'_{3312}$), and (Eq. 1.2.4 in \cite{hearmon})
\begin{equation}
S'_{i,j}= \frac{1}{2} \left( \frac{\partial u_i}{\partial x_j} + \frac{\partial u_j}{\partial x_i} \right)
\end{equation}
with $u_i$ the displacements (elongations) and $x_i$ the coordinates along the three spatial dimensions $i=1,2,3$. 

Then, Eq. \ref{hooke2} can be expanded as: 
\begin{eqnarray}
\label{hooke3}
\frac{\partial u_1}{\partial x_1} = s_{11} \sigma'_{11} + s_{12} \sigma'_{22} + s_{13} \sigma'_{33} 
    + s_{14} \sigma'_{23} + s_{15} \sigma'_{13} + s_{16} \sigma'_{12} 
\nonumber \\
\frac{\partial u_2}{\partial x_2} = s_{21} \sigma'_{11} + s_{22} \sigma'_{22} + s_{23} \sigma'_{33} 
    + s_{24} \sigma'_{23} + s_{25} \sigma'_{13} + s_{26} \sigma'_{12} 
\nonumber \\
\frac{\partial u_3}{\partial x_3} = s_{31} \sigma'_{11} + s_{32} \sigma'_{22} + s_{33} \sigma'_{33} 
    + s_{34} \sigma'_{23} + s_{35} \sigma'_{13} + s_{36} \sigma'_{12} 
\nonumber \\
\frac{\partial u_2}{\partial x_3} + \frac{\partial u_3}{\partial x_2}  = 
   s_{41} \sigma'_{11} + s_{42} \sigma'_{22} + s_{43} \sigma'_{33} 
    + s_{44} \sigma'_{23} + s_{45} \sigma'_{13} + s_{46} \sigma'_{12} 
          \\
\frac{\partial u_1}{\partial x_3} + \frac{\partial u_3}{\partial x_1}  = 
   s_{51} \sigma'_{11} + s_{52} \sigma'_{22} + s_{53} \sigma'_{33} 
    + s_{54} \sigma'_{23} + s_{55} \sigma'_{13} + s_{56} \sigma'_{12} 
\nonumber \\
\frac{\partial u_1}{\partial x_2} + \frac{\partial u_2}{\partial x_1} = 
   s_{61} \sigma'_{11} + s_{62} \sigma'_{22} + s_{63} \sigma'_{33}
    + s_{64} \sigma'_{23} + s_{65} \sigma'_{13} + s_{66} \sigma'_{12} 
\nonumber
\end{eqnarray}

\begin{figure}
\label{fig_moments}
\caption{Double moment bending of a rectangular plate of thickness $T$ 
}
\includegraphics[keepaspectratio,angle=-90,origin=c,width=3.0in]{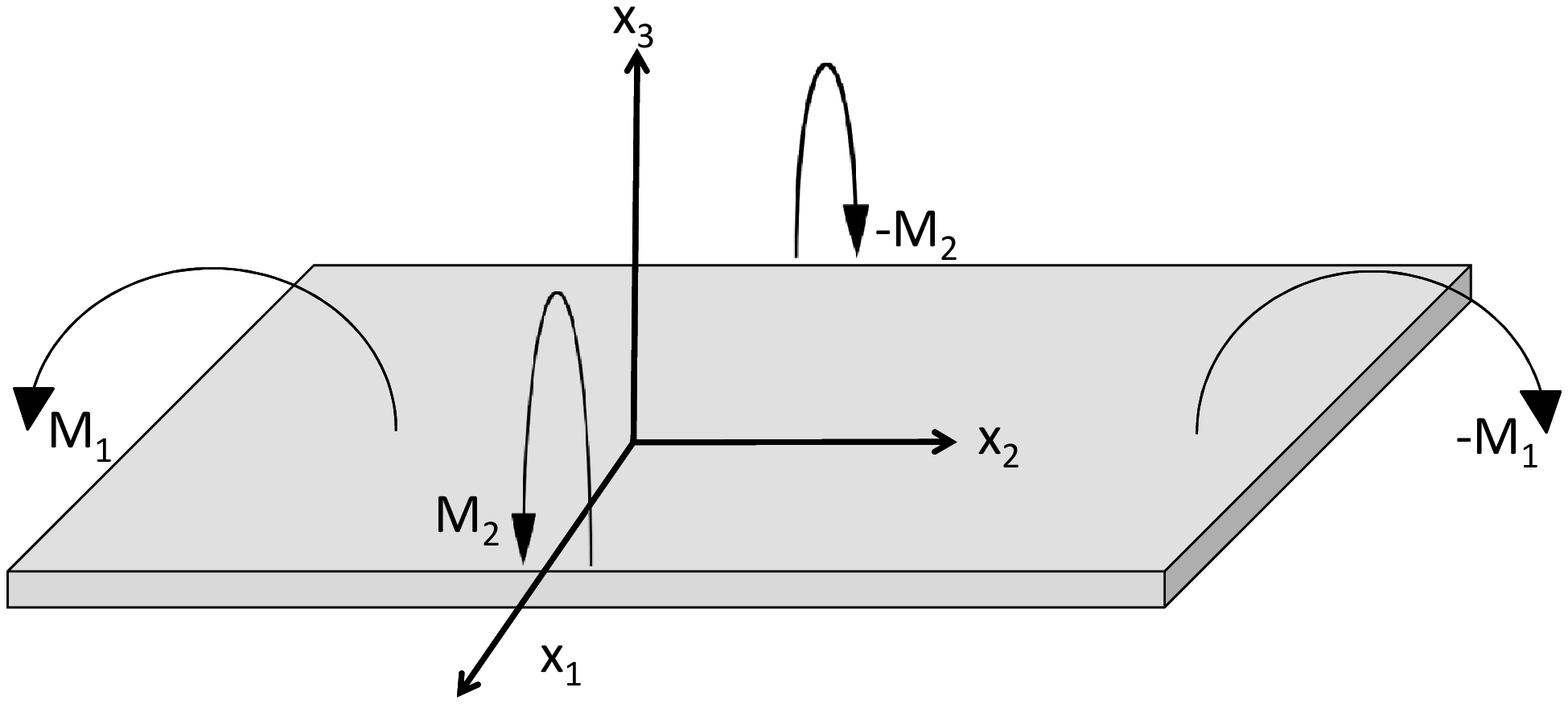}
\end{figure}

In general, the stress tensor is proportional to the generalized torques $M$ (per unit of length) and forces $A$. 
Let us suppose a crystal as a rectangular plate of thickness $T$ with axes 1 and 2 parallel to the the edges, and
axis 3 perpendicular to the surface.
Considering only two torques $M_1$ and $M_2$ (dimensions torque per length), applied in-plane (Fig. 
\ref{fig_moments}), we have:
\begin{gather}
\label{bent222}
\sigma'_{11}= \frac{M_1 x_3}{I} \nonumber \\
\sigma'_{22}= \frac{M_2 x_3}{I} \nonumber \\
\sigma'_{33}= 0 \\ 
\sigma'_{13}=\sigma'_{23}=\sigma'_{12}=0 \nonumber
\end{gather}
where the zero terms are a consequence of considering pure bending, and $I=T^3/12$ is the inertia moment.

Replacing all $\sigma'_{ij}$ into Eq. (\ref{hooke3}) we obtain: 
\begin{eqnarray}
\label{hooke4}
\frac{\partial u_1}{\partial x_1} = \frac{1}{I} \left( s_{11} M_1 x_3 + s_{12} M_2 x_3 \right) \nonumber \\
\frac{\partial u_2}{\partial x_2} = \frac{1}{I} \left( s_{21} M_1 x_3 + s_{22} M_2 x_3 \right) \nonumber \\
\frac{\partial u_3}{\partial x_3} = \frac{1}{I} \left( s_{31} M_1 x_3 + s_{32} M_2 x_3 \right)           \\
\frac{\partial u_2}{\partial x_3} + \frac{\partial u_3}{\partial x_2} = 
   \frac{1}{I} \left( s_{41} M_1 x_3  + s_{42} M_2 x_3  \right)                                \nonumber \\
\frac{\partial u_1}{\partial x_3} + \frac{\partial u_3}{\partial x_1}  = 
   \frac{1}{I} \left( s_{51} M_1 x_3  + s_{52} M_2 x_3 \right)                                 \nonumber \\
\frac{\partial u_1}{\partial x_2} + \frac{\partial u_2}{\partial x_1} = 
   \frac{1}{I} \left( s_{61} M_1 x_3  + s_{62} M_2 x_3 \right)
\nonumber
\end{eqnarray}

Integrating these equation we obtain  (Eq. 4 in \cite{chukhovskii1992}): 

\begin{gather} 
\label{us_integrated}
u_1 = \frac{1}{I} [ (s_{11} M_1 + s_{12} M_2 ) x_1 x_3 \nonumber \\
      +(s_{51} M_1 + s_{52} M_2 ) x_3^2 / 2 
      +(s_{61} M_1 + s_{62} M_2 ) x_2 x_3/2 ]
\nonumber \\
u_2 = \frac{1}{I} [ (s_{21} M_1 + s_{22} M_2 ) x_2 x_3  \nonumber \\ 
      +(s_{41} M_1 + s_{42} M_2 ) x_3^2 / 2 
      +(s_{61} M_1 + s_{62} M_2 ) x_1 x_3/2  ] 
\\
u_3 = \frac{1}{2I} [ -(s_{11} M_1 + s_{12} M_2 ) x_1^2 
      -(s_{21} M_1 + s_{22} M_2 ) x_2^2 \nonumber \\
      -(s_{61} M_1 + s_{62} M_2 ) x_1 x_2 
      +(s_{31} M_1 + s_{32} M_2 ) x_3^2 ] \nonumber
\nonumber
\end{gather}

Replacing $x_3=-T/2$ we obtain the equation of the plate surface  (Eq. 5a in \cite{chukhovskii1992})): 
\begin{gather}
u_3 = -(s_{11} M_1+s_{12} M_2) \frac{x_1^2}{2I}
      -(s_{21} M_1+s_{22} M_2) \frac{x_2^2}{2I} \nonumber \\
      -(s_{61} M_1+s_{62} M_2) \frac{x_1 x_2}{2I}
      +(s_{31} M_1+s_{32} M_2) \frac{T^2}{8I}
\end{gather}

From here, one can obtain the profiles along the $i=1,2$ directions, which are approximately circular
with the form $u_3 = -x_i^2/(2R_i)$. The radii are: (Eq. 5b in \cite{chukhovskii1992})): 
\begin{eqnarray}
\frac{1}{R_1}= s_{11} \frac{M_1}{I}+s_{12} \frac{M_2}{I} \nonumber \\
\frac{1}{R_2}=s_{21} \frac{M_1}{I}+s_{22} \frac{M_2}{I}
\label{radii}
\end{eqnarray}

From Eq.~(\ref{radii}) is possible to calculate the applied torques as a function of the curvatures:
\begin{eqnarray}\label{torques}
\frac{M_1}{I} = \frac{1}{s_{12} s_{21}-s_{11} s_{22}} \left( \frac{s_{12}}{R_2}-\frac{s_{22}}{R_1} \right) \\ \nonumber
\frac{M_2}{I} = \frac{1}{s_{12} s_{21}-s_{11} s_{22}} \left( \frac{s_{21}}{R_1}-\frac{s_{11}}{R_2} \right)
\end{eqnarray}

The Eq. B.8  give the displacements $u_i$ knowing the elastic compliance 
tensor and the torques applied (or more interestingly, via the bending radii using Eq.~\ref{radii}).

\subsection{The compliance tensor for crystals cut along given directions}

The number of independent components in the compliance tensor depends on the crystal symmetry. For a crystal cut along the 
crystallographic axes, triclinic crystals have all 21 components independent, monoclinic crystals have 13, orthorhombic 
crystal 9, tetragonal crystals 6 or 7 and cubic crystals 3 (see table 5 in \cite{hearmon}). An isotropic material has only two 
independent components. For a crystal cut along given directions, fourth-rank compliance tensor must be transformed 
following (Eq. 1.5.2 in \cite{hearmon}):

\begin{equation}
\label{transform}
 s^{''}_{ijkl} = a_{im} a_{jn} a_{ko} a_{lp} s'_{mnop}
\end{equation}
where the elements of the $a$ tensor are the direction cosines of the new axes. Software codes are available 
\cite{VeijoPC} \cite{Schulze}  to transform a generic compliance tensor expressed in the reference frame coincident with the 
crystallographic axes, to a new one where the crystal has been cut following known planes expressed by their Miller indices. 
 
For a cubic crystal cut along its crystallographic axes, the compliance tensor has only three different values. The non-zero
elements are: $s_{12}=s_{21}=s_{13}=s_{31}=s_{23}=s_{32}$, $s_{11}=s_{22}=s_{33}$ and $s_{44}=s_{55}=s_{66}$. Because of 
cubic symmetry, this tensor is independent on the choice of the axes in the crystal (in plane or normal). The components 
of the compliance tensor for the most usual crystals are expressed in Table \ref{table_s}. The components of the compliance 
tensor for a cubic crystal cut along a given direction can be calculated using the generic Eq. (\ref{transform}),
resulting in analytical expressions given by \cite{wortman}. A convenient form easy to implement in computer languages 
is given in \cite{linzhang} \cite{zhangJSR} .

In the particular case of one single torque ($M_2=0$) Eqs. (\ref{radii}) give the main bending radius $R_1=I/(s_{11}M_1)$ and a 
curvature radius in the perpendicular direction $R_2=R_1 (s_{11}/s_{12})$, the anticlastic curvature. The ratio of the 
transverse component of the compliance tensor (in this case along the direction $12$) over the longitudinal component is 
the Poisson's ratio $\nu_{12}=-s_{12}/s_{11}$. For example, a silicon crystal cut along the crystallographic axes has (see 
table \ref{table_s}) $\nu_{12}=-(-2.14)/7.68=0.28$ thus the anticlastic radius is 3.6 times larger than the main
bending radius with the curvature in opposite direction. Note that the Poisson's ratio and the anticlastic radius in a crystal 
cut along directions different from the crystallographic axes are different. 
\newline

\begin{table}
\label{table_s}
\caption{Compliance tensor $s$ elements ($\times 10^{-12}~m^2/N$) for most used cubic perfect crystals 
(Si, Ge: \cite{wortman}, Diamond: \cite{berman}; see review in \cite{Khounsary}). 
For completeness, the relationships between the compliance $s$ and the stiffness $c$ tensors for cubic crystals are included.
}
\vspace{0.3cm}
\begin{tabular}{lccccl}      
            & Si        & Ge       & Diamond   &  &$s$ and $c$ relationships\\
\hline
 $s_{11}$   &  7.68      & 9.64       & 1.04      &  & $(c_{11}+c_{12})/[(c_{11}-c_{12})(c_{11}+2c_{12})]$  \\
 $s_{12}$   & -2.14      & -2.60      & -0.211     &  & $-c_{12}/[(c_{11}-c_{12})(c_{11}+2 c_{12})]$  \\
 $s_{44}$   & 12.6       & 14.9       & 1.93      &  & $1/c_{44}$ \\
 $c_{12}$   &           &           &          &  & $-s_{12}/[(s_{11}-s_{12})(s_{11}+2 s_{12})]$ \\
 $c_{11}$   &           &           &          &  & $(s_{11}+s_{12})/[(s_{11}-s_{12})(s_{11}+2 s_{12})]$ \\
 $c_{44}$   &           &           &          &  & $1/s_{44}$ \\
\end{tabular}
\end{table}

\begin{table}
\label{equivalence}
\caption{Equivalence for the compliance tensor indices for different reference frames used 
in literature:Shi \cite{xianbo_spie}, Honkim{\"{a}}ki \cite{VeijoPC}, Schulze \cite{SchulzeThesis} \cite{Schulze}, 
and Zhang \cite{zhangJSR}.}
\begin{tabular}{lcccc}      
This Paper & Shi  & Honkim{\"{a}}ki & Schulze & Zhang \\
\hline
1 &  3 &  1 & 2  & 2 \\
2 & -2 & -3 & 1  & 3 \\
3 &  1 &  2 & -3 & 1 \\
4 &  6 &  4 & 5  & 5 \\
5 &  5 &  6 & 4  & 6 \\
6 &  4 &  5 & 6  & 4 \\
\end{tabular}
\end{table}

\section{The parameter $c$ in the multilamellar model.}
\label{appendix:ml}
\label{appendix:ml_taupin}
In the multilamellar model, for diffraction occurring in the $x_2,x_3$ plane, the bent crystal is approximated by a stack of 
lamellae which gradually change direction and lattice spacing. 

The reduced curvature $c$ can be expressed as: 

\begin{equation}
\label{c_ml_def}
c = \frac{d\eta}{d A }=
\frac{d\eta}{d \alpha_Z} 
\frac{d\alpha_Z}{d t} 
\frac{d t}{d A}=
\frac{b \Lambda}{\sqrt{|b|} P |\Psi_H|}  \frac{d\alpha_Z}{d t}
\end{equation}
where the definitions of the $\eta$ and $A$ in \cite{Zachariasen} have been used, and $t=x_3$.

Taupin \cite{TaupinThesis} discussed the variation of $\alpha_Z$ as a function of the crystal
deformation and form of the incident wave, which can be chosen in such a way that $\alpha_Z$ is only 
dependent on the thickness direction $t=x_3$. In this case, he found a tensorial expression (\cite{TaupinThesis} Eq. II.1.7):

\begin{equation}
\frac{d\alpha_Z}{d t}=-\frac{2}{\gamma_0}
\sum_{k=1}^3\sum_{j=1}^3\sum_{i=1}^3 V^{H}_{k}V^{0}_{j}(V^H_i-V^0_i)\frac{\partial^{2} u_i}{\partial x_j \partial x_k},
\end{equation}
where $\vec{V}^{0,H}=(V_1^{0,H},V_2^{0,H},V_3^{0,H})$ are unitary vectors along the incident and diffraction
directions, as defined in the text. 

Performing the summation in the diffraction plane $23$ (thus $V^{H}_{1}=V^{0}_{1}=0$) one gets:
 
\begin{gather}
\frac{d\alpha_Z}{d t}=-\frac{2}{\gamma_0} [
V^{H}_{2} V^{0}_{2}(V^H_3-V^0_3)\frac{\partial^{2} u_3}{\partial^{2} x_2}+ \nonumber \\
V^{H}_{2} V^{0}_{3}(V^H_2-V^0_2)\frac{\partial^{2} u_2}{\partial x_2\partial x_3} + \nonumber \\
V^{H}_{3} V^{0}_{3}(V^H_2-V^0_2)\frac{\partial^{2} u_2}{\partial x_3^{2}}+ \\
V^{H}_{3} V^{0}_{3}(V^H_3-V^0_3)\frac{\partial^{2} u_3}{\partial x_3^{2}}+ \nonumber \\
V^{H}_{3} V^{0}_{2}(V^H_2-V^0_2)\frac{\partial^{2} u_2}{\partial x_2\partial x_3} \nonumber
] \nonumber
\end{gather}

Inserting the derivatives calculated from the equations for the displacements B.8, 
one obtains:

\begin{gather}
\label{alphazovert}
\frac{d\alpha_z}{d t}=-\frac{2}{\gamma_0} [  
(s_{21}\frac{M_1}{I} + s_{22}\frac{M_2}{I})  A_1+ \\
(s_{31}\frac{M_1}{I} + s_{32}\frac{M_2}{I})  A_2+
(s_{41}\frac{M_1}{I} + s_{42}\frac{M_2}{I})  A_3 \nonumber
]
\end{gather}

with: 

\begin{gather}
A_1 = -V^H_2 V^0_2 (V^H_3-V^0_3) + V^H_2 V^0_3 (V^H_2-V^0_2)  + V^H_3 V^0_2 (V^H_2-V^0_2) = \nonumber \\
(\gamma_0-\gamma_H)(1+\gamma_0 \gamma_H) \nonumber \\
A_2 =  V^H_3 V^0_3 (V^H_3-V^0_3) = \gamma_0 \gamma_H (\gamma_H - \gamma_0) \nonumber \\
A_3 =  V^H_3 V^0_3 (V^H_2-V^0_2) = \nonumber \\
              \gamma_0 \gamma_H (\sqrt{1-\gamma_H^2} - \sqrt{1-\gamma_0^2}) 
\end{gather}

For computing $A_3$ it is preferred to use  the form without the square toot, to avoid incertitude due to the double sign.
For particular case widely treated in literature of meridionally bending $M_1=0$, $M_2/I=1/(R_2 s_{22})$ and isotropic 
(Poisson's ratio $\nu=-s_{23}/s_{22}$, and $s_{42}=0$) we have: 

\begin{gather}
\frac{d\alpha_z}{d t}=-\frac{2}{\gamma_0 R_2} [A_1-\nu A_2] 
\end{gather}

which gives a $c$ parameter like \cite{Caciuffo}:

\begin{equation}
c = \frac{\lambda (b-1) |\gamma_H|} {\pi P^2 |\Psi_H|^2  R_2} 
\left[ 1+b(1+\nu) \gamma_H^2 \right],
\end{equation}

For the particular case of sagittal bending, $M_2=0$ and  $M_1/I=1/(R_1 s_{11})$  the Eq.~\ref{alphazovert} reduces to: 
\begin{equation}
\frac{d\alpha_Z}{d t}=-\frac{2}{\gamma_0 R_1}
(\frac{s_{21}}{s_{11}} A_1 + \frac{s_{31}}{s_{11}} A_2 + \frac{s_{41}}{s_{11}} A_3)
\end{equation}

therefore: 
\begin{equation}
c = - \frac{|\gamma_h|}{\gamma_h} \frac{\lambda}{P^2 |\Psi_H|^2 R_1} \frac{1}{\pi}
(\frac{s_{21}}{s_{11}} A_1 + \frac{s_{31}}{s_{11}} A_2 + \frac{s_{41}}{s_{11}} A_3)
\end{equation}



\section{The parameter $\beta$ in the PP model}
\label{appendix:pp}

The strain gradient $\beta$ of an asymmetric Laue crystal with respect to the crystal surface is defined by 
\begin{equation}
\label{beta_xianbo}
\beta = \frac{1}{P |\Psi_{{H}}|}     \frac{1}{k \sqrt{|\gamma_0 \gamma_H|} }  \mathbf{G}
\end{equation}

where {\bf G} is defined as:
\begin{equation}
\label{G}
\mathbf{G} \equiv
\frac{\partial^2 (\vec{H}.\vec{u})}{\partial V_0 \partial V_H}
\end{equation}

For calculating $\mathbf{G}$ we restrict the calculation to the diffraction plane $23$, thus replacing 
$\vec{H}.\vec{u}=(n^H_2 u_2+n^H_3 u_3)/d_{hkl}$, $\vec{V}_0=(0,V^0_2,V^0_3)$ and $\vec{V}_H=(0,V^H_2,V^H_3)$
we obtain:

\begin{gather}
\mathbf{G} = (\vec{V}_0 . \vec{\nabla}_r) (\vec{V}_H . \vec{\nabla}_r) (\vec{H} . \vec{u}) =  \nonumber \\
\frac{1}{d_{hkl}} [
V^{0}_{3}V^{H}_{3}n^H_3\left(  \frac{M_2}{I}s_{23}   +  \frac{M_1}{I}s_{13}   \right)+ 
V^{0}_{3}V^{H}_{3}n^H_2\left(  \frac{M_2}{I}s_{24}   +  \frac{M_1}{I}s_{14}   \right)+ \nonumber \\
\left(V^{0}_{2}V^{H}_{3}n^H_2+V^{0}_{3}V^{H}_{2}n^H_2-V^{0}_{2}V^{H}_{2}n^H_3\right) 
\left(   \frac{M_2}{I}s_{22}  +   \frac{M_1}{I}s_{12}   \right) 
] 
 \equiv \nonumber \\
\frac{1}{d_{hkl}} [ \mathbf{G_1} \left(     \frac{M_2}{I}s_{23}  +    \frac{M_1}{I}s_{13}   \right)+ 
                       \mathbf{G_2} \left(  \frac{M_2}{I}s_{24}  +    \frac{M_1}{I}s_{14}   \right)+ \nonumber \\
                       \mathbf{G_3} \left(  \frac{M_2}{I}s_{22}  +    \frac{M_1}{I}s_{12}   \right) 
] 
\label{veijo} 
\end{gather}

Considering that $\gamma_{0,H}=V^{0,H}_3$, and replacing the vector components by their angular
expressions (Eq. \ref{vector2angles}) we get:

\begin{gather}
\label{gs}
\mathbf{G_1} =V^{0}_{3}V^{H}_{3}n^H_3 =  \gamma_0 \gamma_H \sin\chi \nonumber \\
\mathbf{G_2} =V^{0}_{3}V^{H}_{3}n^H_2 = \gamma_0 \gamma_H \cos\chi  \\
\mathbf{G_3} = V^{0}_{2}V^{H}_{3}n^H_2+V^{0}_{3}V^{H}_{2}n^H_2-V^{0}_{2}V^{H}_{2}n^H_3 = \nonumber \\ 
-(1+\gamma_0 \gamma_H) \sin\chi =  -\left( 1+\frac{\cos2\theta_B + \cos2\chi}{2} \right)  \sin\chi  \nonumber 
\end{gather}

For bending the crystal only in meridional direction with a curvature radius $R_m\equiv R_2$ we have $M_1=0$ 
and $M_2/I=(R_2 s_{22})^{-1}$ (from Eq. (\ref{radii})). Inserting these values into Eq. (\ref{veijo}) we obtain
(note also that  $2\gamma_0 \gamma_H = \cos2\theta_B+\cos2\chi$): 

\begin{equation}
\label{g_clemens}
\mathbf{G} = \frac{-1}{d_{hkl} R_2} \sin\chi 
\left[ 1+ \frac{\cos2\theta_B + \cos2 \chi}{2}\left( 1-\frac{s_{23}+\cot\chi s_{24}}{s_{22}}\right)\right]
\end{equation}
which inserted in (\ref{beta_xianbo}) gives:
\begin{eqnarray}
\label{beta_clemens}
\beta = -\frac{2 \sin\chi \tan\theta_B}{P \sqrt{\Psi_H\Psi_{\bar{H}}}} \frac{1}{R_2}  \times \nonumber \\
\left[ 1+ \frac{\cos2\theta_B + \cos2 \chi}{2}\left( 1-\frac{s_{23}+\cot\chi s_{24}}{s_{22}}\right)\right]
\end{eqnarray}
The $\beta$ value in Eq. (\ref{beta_clemens}) was first obtained by \cite{Schulze}. The equations are identical 
considering the different choice of the axis. 
Noticeably, there are typos in the indices of the compliance 
tensor in Eq.~5 of \cite{Schulze} (as well as in Eq.~(37) of \cite{SchulzeThesis}). If using their reference system 
as defined in Fig. 1 of \cite{Schulze} (or Fig.~7 in \cite{SchulzeThesis}) one should obtain 
an equation like Eq.~(\ref{beta_clemens}) but replacing $s_{22}\to s_{11}, s_{23}\to s_{13}, s_{24}\to s_{15}$ 
(see Table~\ref{equivalence}). The equation as written in these two references correspond to a different reference
frame, like the one shown in Fig.~10 of \cite{SchulzeThesis}.

The case of isotropic materials can be obtained setting $s_{24}=0$ in Eq. (\ref{beta_clemens}) and considering 
the definition of Poisson's ratio, as minus the ratio of the compliance term element along the direction transversal 
to the curvature and the one along the curved direction  (i.e., $\nu=-s_{23}/s_{22}$. We obtain \cite{Sanchez1997}:

\begin{equation}
\label{beta_isotropic}
\beta = -\frac{2 \sin\chi \tan\theta_B}{P \sqrt{\Psi_H\Psi_{\bar{H}}}} \frac{1}{R_2} 
\left[ 1+ \frac{\cos2\theta_B + \cos2 \chi}{2} (1+\nu) \right]
\end{equation}

The case of single bending in sagittal direction was studied in \cite{xianbo_spie}. Setting 
$R_s\equiv R_1$, $M_2=0$ and $M_1/I=(R_1 s_{11})^{-1}$ (from Eq. (\ref{radii})) one obtains:
 
\begin{equation}
\label{g_xianbo}
\mathbf{G} = \frac{-1}{d_{hkl} R_1} \sin\chi 
\left[ \frac{s_{12}}{s_{11}} + \gamma_0\gamma_H \left( \frac{s_{12}}{s_{11}}-\frac{s_{13}+\cot\chi s_{14}}{s_{11}}\right)\right]
\end{equation}
which inserted in (\ref{beta_xianbo}) gives:
\begin{eqnarray}
\label{beta_xianbo}
\beta = -\frac{2 \sin\theta_B}{P |\Psi_H| \sqrt{|\gamma_0 \gamma_H|} } \frac{1}{R_1}\sin\chi \times \nonumber \\
\left[ 
-\gamma_0\gamma_H \frac{s_{13}+\cot\chi s_{14}}{s_{11}}
+(1+\gamma_0\gamma_H)\frac{s_{12}}{s_{11}}
\right]
\end{eqnarray}

which corresponds to Eq. 18 in \cite{xianbo_spie}.

\bibliography{crystal_bent}   
\bibliographystyle{iucr}   

\end{document}